\documentclass[aps,pra,showpacs,twoside,10pt,floatfix,nofootinbib,longbibliography,twocolumn]{revtex4-2}
\usepackage[colorlinks=true, citecolor=red, urlcolor=blue ]{hyperref}
\usepackage{epsfig,newlfont,amssymb,amsfonts,amsmath,bm,subfigure,palatino,mathtools,amsthm,braket,soul,enumitem,color,times,comment,geometry,xcolor,contour}
\usepackage[normalem]{ulem}
\definecolor{defcolor}{HTML}{59368F}

\definecolor{subdefcolor}{HTML}{47418C}

\newcommand{\ketbra}[2]{|#1\rangle\langle#2|}

\begin{document}

\title{Robust quantum metrology using disordered probes}
\author{Vishnupriya K., Harikrishnan K. J., and Amit Kumar Pal}
\affiliation{Department of Physics, Indian Institute of Technology Palakkad, Palakkad 678 623, India}
\date{\today}

\begin{abstract}
Disorder is ubiquitous in quantum devices including quantum probes designed and fabricated for quantum parameter estimation and sensing. We investigate the robustness of a quantum probe against the presence of glassy disorder. We define a disorder marker quantifying the effect of the disorder by expanding the quantum Fisher information in terms of different orders of the standardized central moments of the disorder-distributions. We classify the quantum probes in terms of the possible values of the disorder marker, and analytically show, for a disorder-sensitive probe with identical and weak disorder on all or a subset of the parameters of the probe-Hamiltonian, that the absolute value of the disorder marker exhibits a quadratic dependence on the disorder strength. We derive a robustness scale intrinsic to the probe that competes with the disorder, and provide a prescription  for estimating the maximum disorder strength that the probe can withstand from the disorder-free probe-Hamiltonian for a given initial state of the probe, which can be computed without the disorder averaging. We also extend the prescription to classical Fisher information, and provide an operational perspective to the proposed marker in experiments. We demonstrate our results in the case of a single-qubit probe under disordered magnetic field, and a multi-qubit probe described by a disordered one-dimensional Kitaev model with nearest-neighbor interactions.  
\end{abstract}

\maketitle

\section{Introduction}
\label{sec:intro}

Quantum metrology and sensing~\cite{Giovannetti2004,Giovannetti2006,Giovannetti2011,Paris2009,Degen2017} have rapidly advanced in the last few years, leading to designs and fabrications of highly sensitive and precise quantum sensors for detecting gravitational~\cite{Peters1999,Fixler2007,El-Neaj2020,Stray2022}, magnetic~\cite{Dang2010,Bal2012,Baumgart2016}, and electric fields~\cite{Bian2021,Qiu2022,JoshuaGordon,Zhang2024,Gilmore2021,Esat2024}, and proposing the use of quantum resources  to surpass the classical limit of sensing~\cite{Giovannetti2004,Giovannetti2006,Giovannetti2011}. Implementation of practical quantum sensors has been possible using ultracold atoms~\cite{Takamoto2005,Hinkley2013,Bloom2014,Baumgart2016,Bongs2019}, trapped ions~\cite{Leibfried2004,Maiwald2009,Biercuk2010,Brownnutt2015},  superconducting circuits~\cite{Danilin2018,Wang2019} and photonic systems~\cite{Holland1993,Mitchell2004,Budker2007,Higgins2007,Pezze2007}, expanding its reach to detecting extremely weak signals corresponding to the gravitational wave~\cite{Schnabel2010,LIGO2011,Danilishin2012,El-Neaj2020,Tse2019} and dark matter~\cite{El-Neaj2020,Jiang2021}. In quantum sensing, typically, a quantum probe state encodes the parameter $\theta$ of a time-independent probe Hamiltonian $H$ via a time-evolution $\ket{\psi_{t}}=U_{t}\ket{\psi}$
with $U_{t}=\exp\left[-\text{i}H t\right]$, with a \emph{pre-decided} initial pure state $\ket{\psi}$.  The quantum Fisher information (QFI)~\cite{Helstrom1967,Helstrom1969,Braunstein1994} 
\begin{eqnarray}
    F= 4\big(\langle\partial_{\theta}\psi_{t}| \partial_{\theta}\psi_{t}\rangle-\left|\langle\psi_{t}|\partial_{\theta}\psi_{t}\rangle \right|^2\big)
\end{eqnarray}
quantifies the sensitivity of $\ket{\psi_t}$ to detect an infinitesimal change in $\theta$,  with $\partial_{\theta}\ket{\psi_{t}}= -\text{i} U_{t} G\ket{\psi}$, where the hermitian operator 
\begin{equation}\label{eq:operator}
G = \int_0^t ds U^\dagger_{s} \partial_{\theta} H U_{s}
\end{equation}
is the QFI generator (QFIG)~\cite{PangBrun2014,Skotiniotis2015}, leading to  $F_\psi = 4 \left(\Delta_\psi G\right)^2$, with $\Delta_\psi G=\sqrt{ \langle\psi|G^2| \psi\rangle - \langle\psi|G|\psi\rangle^2}$. For a fixed $H$, the initial state $\ket{\psi}$ resulting in the maximum QFI $F_{\psi_m}$ is $\ket{\psi_m}$, with   $\ket{\psi_m}=\left(\ket{\lambda^{\min}}+\text{e}^{\text{i}\beta}\ket{\lambda^{\max}}\right)/\sqrt{2}$~\cite{Giovannetti2006,Skotiniotis2015}, where  
$\ket{\lambda^{\max}}$ $\left(\ket{\lambda^{\min}}\right)$ is the eigenvector  of $G$ corresponding to the maximum (minimum) eigenvalue $\lambda^{\max}$ ($\lambda^{\min}$).

However, imperfections in real-world quantum measurement devices can significantly affect performance~\cite{Clerk2010,Jiao2025}. If the probe is \emph{open} to environmental fluctuations, parameter encoding is governed by a dissipative quantum master equation rather than unitary evolution~\cite{Demkowicz2012,Kolodynski2013,liu2014,liu2015,Xiao2022,berrada2025}. Moreover, the post-encoding measurement can be noisy~\cite{Zhou2023}, and the probe state itself can be imperfect due to decoherence, thermalization, or inefficient initialization~\cite{Wasilewski2010,Datta2011,Escher2011,Yu2013,Demkowicz2014,Haine2015}. 

Apart from environmental interference and measurement errors, imperfections such as \emph{glassy} or \emph{quenched disorder}~\cite{Sherrington1975,Edwards1975,Derrida1981} are ubiquitous in quantum sensing devices. These systems often contain random interactions, impurities, or spatial inhomogeneities, leading to a quenched distribution of local fields and couplings that remains effectively static during time evolution. Glassy disorder can arise in several quantum metrology platforms, including spin-ice magnetometry~\cite{Kirschner2018}, with intrinsic glassiness~\cite{Samarakoon2022,Billington2025,Zapke2025} (c.f.~\cite{Freedberg2024}), Rydberg-atom platforms~\cite{Dietsche2019,Zou2023} with  experimental evidence of disorder~\cite{Lesanovsky2013,PerezEspigares2018,Signoles2021,JuliaFarre2024,Brodoloni2025}, and Ramsey-interferometry-based atomic clocks~\cite{Ramsey1991,Clairon1995,Wynands2005,Takamoto2005,Ludlow2008,Heavner2014,Peil2014,Ludlow2015,Guena2015,Kruse2016,Olson2019,Bregazzi2024,Fartmann2025} (c.f.~\cite{Ramsey1950}), where temporally correlated noise can generate long-time correlations that effectively mimic glassy disorder~\cite{Dorscher2020}.

Despite clear motivations, systematic investigation of the effect of the presence of disorder in the probe Hamiltonian is scarce in literature, and only a few  studies phenomenologically addressing the problem exist. Mothsara et. al.~\cite{Mothsara2025} proposed a probe modeled by a variable-range many-body quantum spin chain with a moderate transverse magnetic field, and demonstrated robustness of its performance against impurities in the
magnetic field, when the interaction is chosen to be of Ising-type. Bhattacharyya et. al.~\cite{Bhattacharyya2024} studied the atomic clock measuring time depending  on the precision measurement of the frequency of a two-level atom, and demonstrated that the accuracy of the measurement of the frequency enhances, as well as the requirement of entanglement content of the initial probe states reduces when disorder is present in the atom. Recently, Burte et. al.~\cite{2026arXiv_Keshav} explored the effect of disorder on the quantum Fisher information in topological systems. While these results strongly motivate the exploration of disorder effects in quantum sensing devices, a comprehensive and quantitative characterization of their impact on metrological performance remains an open challenge.

Towards addressing this, we consider a quantum probe described by a Hamiltonian whose parameters are suffering from glassy disorder. We observe that the strength of the disorder in a specific parameter corresponds to the standard deviation of the disorder distribution. This motivates us to expand the quenched averaged QFI computed for the disordered quantum probe with respect to a specific choice of the initial state of the probe in terms of the standardized central moments of the disorder distributions, where the zeroth order term is identified as the QFI of the clean probe. Using this expansion, we define a \emph{disorder marker} to quantitatively investigate the effect of glassy disorder on QFI in a disordered probe. 

Using the possible values of the disorder marker, we classify the disordered probe into three possible categories, namely, the \emph{disorder-immune}, the \emph{disorder-sensitive}, and the \emph{disorder-enhanced} probe. In this paper, we specifically focus on the disorder-sensitive probe, for which the absolute value of the disorder marker varies between zero and one. We prove that in the case of identical disorder with same disorder strength $\sigma$ acting on all disordered parameters, the absolute value of the disorder marker exhibits a quadratic variation with $\sigma$ when $\sigma\ll\sigma^{\max}$, which represents the maximum disorder strength the probe can withstand, at which the absolute value of the disorder marker saturates to $1$. We further show that $\sigma^{\max}$ is a function of solely the clean-probe Hamiltonian, and the chosen initial state of the probe, thereby estimating the robustness of a given quantum probe via the QFI possible without performing disorder averaging of the QFI. It is worthwhile to  point out here that this clearly differentiates between our approach, and the avenues of performing the disorder-averaging of relevant observables followed predominantly in literature~\cite{Doty1992,Tang2015,Mulanix2019,Park2021,Erpelding2026}.

In practice, an experiment with the designed probe involves a projective measurement producing a random outcome drawn from a probability distribution specific to the state of the probe and the choice of the measurement. The uncertainty in estimating the parameter under focus from these outcomes depends on the classical Fisher information (CFI)~\cite{Braunstein1994}. We extend our analysis to this framework also, and derive the disorder marker, leading to a definition of the maximum sustainable disorder strength. We assume the realistic scenario where the experimentalist does not have control on the disorder, and chooses a fixed disorder-ignorant measurement. Defining a set of clean optimal measurement that leads to the same quantum and classical Fisher information in the disorder-free scenario,  we establish a specific hierarchy between the maximum sustainable disorder strengths predicted from the quantum and the classical Fisher information.

We demonstrate our results with  two specific examples. We first consider a single-qubit probe under a disordered magnetic field, and show that the robustness of the probe is maximum for a  specific choice of state among the class of quantum states maximizing the QFI for the clean single-qubit probe. We also demonstrate the predictions corresponding to the experimental scenario where a fixed measurement is chosen irrespective of the disorder realization. Further, motivated by the current interest in quantum metrology using many-body systems~\cite{Paris2016,Hauke2016,Frerot2018,Ilias2022,Ding2022,Angelo2022,DiCandia2023,Montenegro2025}, we consider a multi-qubit quantum probe described by a one-dimensional (1D) Kitaev model with nearest-neighbor (NN) interactions~\cite{Kitaev2001} as the second example, and analytically estimate $\sigma^{\max}$ as a function of the relevant system parameters.

\begin{figure}
    \centering
    \includegraphics[width=0.8\linewidth]{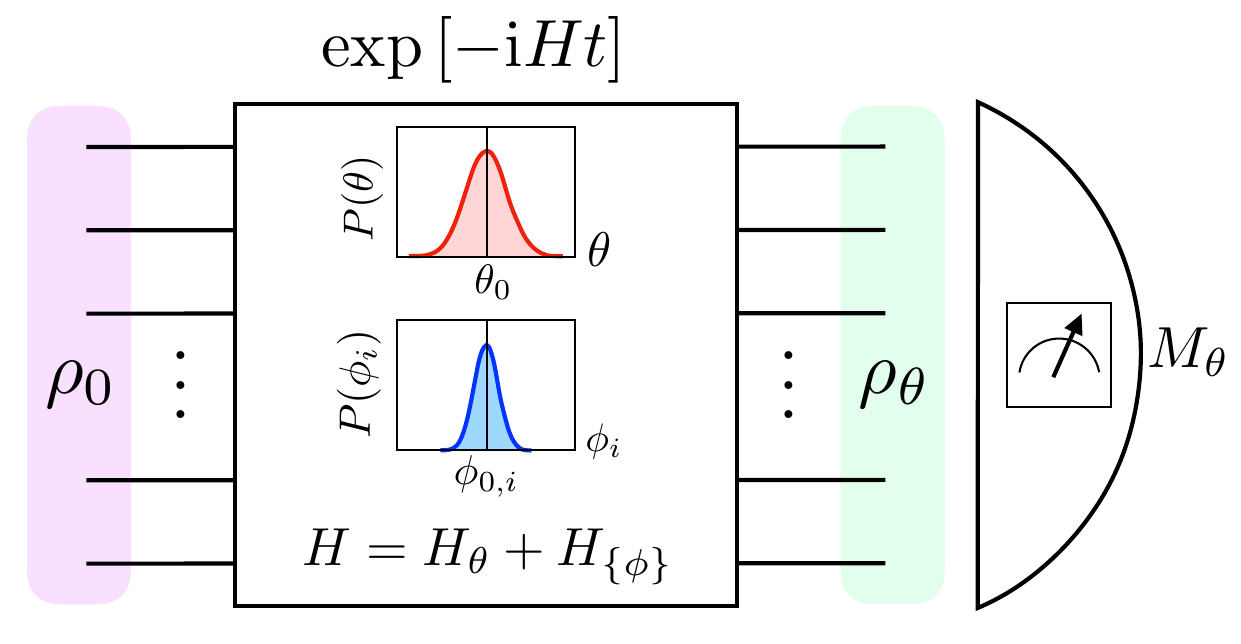}
    \caption{\textbf{Parameter estimation with a disordered probe.} The probe Hamiltonian, \(H=H_\theta+H_{\{\phi\}}\), depends on parameters \(\theta\) and \(\{\phi\}\), each subject to glassy disorder. For example, \(\theta\) is drawn from a probability distribution with mean \(\theta_0\) and standard deviation \(\sigma\). The parameter is encoded through the evolution \(\rho_0\to\rho_\theta=U\rho_0U^\dagger\), where \(U=\exp[-\mathrm{i}Ht]\), followed by a measurement of \(\theta_0\).
    }
    \label{fig:schematics}
\end{figure}

The rest of the paper is organized as follows. In Sec.~\ref{sec:methodology}, we  expand the QFI of a disordered quantum probe in terms of the standardized central moments of the disorder distribution,  and introduce the definition of the disorder marker and the corresponding  classification of the disordered probes. In Sec.~\ref{sec:operational_perspective}, we elaborate on the implications of the disorder marker in a realistic experimental scenario, where the experimentalist has no access to the information regarding the disorder realization, and a fixed measurement is chosen for all disorder realizations. The demonstration of these results in the case of the single-qubit quantum probe under a disordered magnetic field is included in  Sec.~\ref{sec:single_qubit}, while the case of a multi-qubit quantum probe described by the 1D Kitaev model with NN interactions is discussed in Sec.~\ref{sec:kitaev_model}. Sec.~\ref{sec:conclude} contains concluding remarks and outlook.

\section{Quenched quantum Fisher information} 
\label{sec:methodology}

For a systematic analysis, consider a generic disordered \(N\)-body Hamiltonian,
\begin{eqnarray}
   H=\sum_{n=1}^N \phi_n H_n, 
\end{eqnarray}
where \(H_n\) acts non-trivially on an \(L_n\)-body subsystem (\(L_n\leq N\)). Each parameter \(\phi_n\) is drawn from a distribution \(P_n(\phi_n)\) with mean \(\phi_{0,n}\), such that \(\phi_n=\phi_{0,n}+\delta\phi_n\), where \(\delta\phi_n\) (\(|\delta\phi_n|\ll |\phi_{0,n}|\)) denotes random fluctuations. Taking \(\phi_{0,1}\equiv\theta\) as the parameter to be estimated (see Fig.~\ref{fig:schematics}), the Hamiltonian can be written as
\begin{eqnarray}
    H=H_\theta+\sum_{n=2}^N \phi_{0,n}H_n+\sum_{n=1}^N \delta\phi_n H_n,
\end{eqnarray}
thereby naturally decomposing \(H\) into a disorder-free \emph{clean probe} (CP) with Hamiltonian \(H_0\), and a disordered contribution \(V\), given by 
\begin{eqnarray}
\label{eq:disorder_induced_part}
H_0&=&H_\theta+\sum_{n=2}^N \phi_{0,n} H_n,\;\;
V=\sum_{n=1}^N \delta\phi_n H_n,
\end{eqnarray}
so that \(H=H_0+V\), with \([H_0,V]\neq 0\) in general. The disorder strength in the \(n\)th term is characterized by the standard deviation \(\sigma_n\) of \(P_n(\phi_n)\). In the vanishing-disorder or the CP limit, \(\sigma_n\to 0\ \forall n\), one recovers \(H=H_0\). We further assume the probe is initialized perfectly in the state \(\ket{\psi}\).

To isolate the effect of disorder, we work in the interaction picture with respect to \(H_0\), and define the interaction-picture propagator \(\mathcal{U}_s=U_{0,s}^\dagger U_s\), with \(U_s=\exp[-\mathrm{i}Hs]\) and \(U_{0,s}=\exp[-\mathrm{i}H_0s]\). Eq.~(\ref{eq:operator}) therefore becomes
\begin{eqnarray}
G=\int_0^t ds\, \mathcal{U}_s^\dagger \mathcal{H}_{\theta,s}\mathcal{U}_s,
\end{eqnarray}
where 
\begin{eqnarray}
    \mathcal{H}_{\theta,s}=U_{0,s}^\dagger \partial_\theta H\, U_{0,s}.
\end{eqnarray}
Expanding \(\mathcal{U}_s\) using the Dyson series~\cite{Dyson1949,SakuraiQM} (see Appendix~\ref{app:dyson_expansion} for details) yields
\begin{eqnarray}
    G=\sum_{k=0}^\infty G^{(k)},
    \label{eq:G_Dyson}
\end{eqnarray}
where the zeroth-order term,
\begin{eqnarray}
    G^{(0)}=\int_0^t ds\, \mathcal{H}_{\theta,s},
\end{eqnarray}
corresponds to the QFIG of the clean system. The first- and second-order corrections are
\begin{eqnarray}
    G^{(1)}&=&\sum_n \delta\phi_n G_n^{(1)},\nonumber\\
    G^{(2)}&=&\sum_{n_1,n_2}\delta\phi_{n_1}\delta\phi_{n_2}G_{n_1n_2}^{(2)},
\end{eqnarray}
with
\begin{eqnarray}
G_n^{(1)}&=&\mathrm{i}\int_0^t ds\int_0^s ds_1
\left[\mathcal{H}_{n,s_1},\mathcal{H}_{\theta,s}\right],\nonumber\\
G_{n_1n_2}^{(2)}
&=&\int_0^t ds
\left(\int_0^s ds_1\, \mathcal{H}_{n_1,s_1}\right)
\mathcal{H}_{\theta,s}
\left(\int_0^s ds_1\, \mathcal{H}_{n_2,s_1}\right)\nonumber\\
&&-\int_0^t ds\int_0^s ds_1\int_0^{s_1} ds_2
\big(
\mathcal{H}_{n_1,s_2}\mathcal{H}_{n_2,s_1}\mathcal{H}_{\theta,s}
\nonumber\\
&&\qquad\qquad\qquad\qquad+\mathcal{H}_{\theta,s}\mathcal{H}_{n_2,s_1}\mathcal{H}_{n_1,s_2}
\big),
\end{eqnarray}
where 
\begin{eqnarray}
    \mathcal{H}_{n,s}=U_{0,s}^\dagger H_n U_{0,s}.
    \label{eq:H_ns}
\end{eqnarray}
Consequently, the QFI variance for a disorder realization \(\{\delta\phi_n\}\) admits the expansion
\begin{eqnarray}\label{eq:qfig_expansion}
\Delta_\psi^2 G
&=&
\Delta^2_\psi G^{(0)}
+\sum_n \delta\phi_n \tilde{G}_n^\psi
\nonumber\\&&+\sum_{n_1,n_2}\delta\phi_{n_1}\delta\phi_{n_2}\tilde{G}_{n_1n_2}^\psi
+\cdots,
\end{eqnarray}
up to second order in fluctuations, where
\begin{eqnarray}\label{eq:G_first_second_order}
\tilde{G}_n^\psi
&=&
\langle G^{(0)}G_n^{(1)}\rangle
+\langle G_n^{(1)}G^{(0)}\rangle
-2\langle G^{(0)}\rangle\langle G_n^{(1)}\rangle,\nonumber\\
\tilde{G}_{n_1n_2}^\psi
&=&
\langle G_{n_1}^{(1)}G_{n_2}^{(1)}\rangle
+\langle G^{(0)}G_{n_1n_2}^{(2)}\rangle
+\langle G_{n_1n_2}^{(2)}G^{(0)}\rangle\nonumber\\
&&
-2\langle G^{(0)}\rangle\langle G_{n_1n_2}^{(2)}\rangle
-\langle G_{n_1}^{(1)}\rangle\langle G_{n_2}^{(1)}\rangle.
\end{eqnarray}
All expectation values are evaluated in the disorder-independent probe state \(\ket{\psi}\) (see Appendix~\ref{app:dyson_expansion} for third-order terms).

A meaningful characterization of an arbitrary quantity \(Q\) for the disordered many-body Hamiltonian \(H\) is obtained through the quenched average~\cite{Nandkishore2015,Abanin2019},
\begin{eqnarray}\label{eq:disorder_average}
\overline{Q}
=
\int_{-\infty}^{\infty}\cdots\int_{-\infty}^{\infty}
\prod_n d\phi_n\,
Q(\{\phi_n\})P(\{\phi_n\}),
\end{eqnarray}
where \(P(\{\phi_n\})\) is the disorder distribution and \(Q(\{\phi_n\})\) denotes the value of \(Q\) for a given disorder realization \(\{\phi_n\}\). We focus on an \emph{initially clean probe}, where the initial probe state is unaffected by disorder. Assuming uncorrelated disorder for different \(n\), i.e., \(P(\{\phi_n\})=\prod_n P_n(\phi_n)\), and noting that the coefficients \(\tilde{G}_{n_1n_2\cdots n_r}\) in Eq.~(\ref{eq:qfig_expansion}) are disorder independent, the quenched variance becomes
\begin{eqnarray}\label{eq:quenched_variance_expansion}
\overline{\Delta_\psi^2 G}
=
\Delta_\psi^2 G^{(0)}
+\sum_{r=2,3,\cdots}\sum_n
q_n^{(r)}\tilde{G}_n^{\psi,(r)}.
\end{eqnarray}
Here,
\(\tilde{G}_n^{\psi,(r)}
=
\tilde{G}_{n_1n_2\cdots n_r}^\psi\)
with \(n_i=n\;\forall i\), and
\begin{eqnarray}
    q_n^{(r)}
=
\int_{-\infty}^{\infty}
(\phi_n-\phi_{0,n})^r
P_n(\phi_n)\, d\phi_n
\end{eqnarray}
is the \(r\)th central moment of \(P_n(\phi_n)\), satisfying \(q_n^{(1)}=0\). Consequently,
\begin{eqnarray}\label{eq:quenched_qfig_expansion}
\overline{F_\psi}
=
F_{0,\psi}
+4\sum_{r=2,3,\cdots}\sum_n
q_n^{(r)}\tilde{G}_n^{\psi,(r)},
\end{eqnarray}
where \(\overline{F_\psi}=4\overline{\Delta_\psi^2 G}\), and \(F_{0,\psi}\) is the Fisher information of the clean system for an arbitrary probe state \(\ket{\psi}\) (see discussion preceding Eq.~(\ref{eq:operator})). We denote by \(F_{0,\psi_m}\) the maximum of \(F_{0,\psi}\) over all initial states, attained at \(\ket{\psi_m}\), and refer to this as the QFI-maximized CP.

\begin{figure*}
    \centering
    \includegraphics[width=0.9\linewidth]{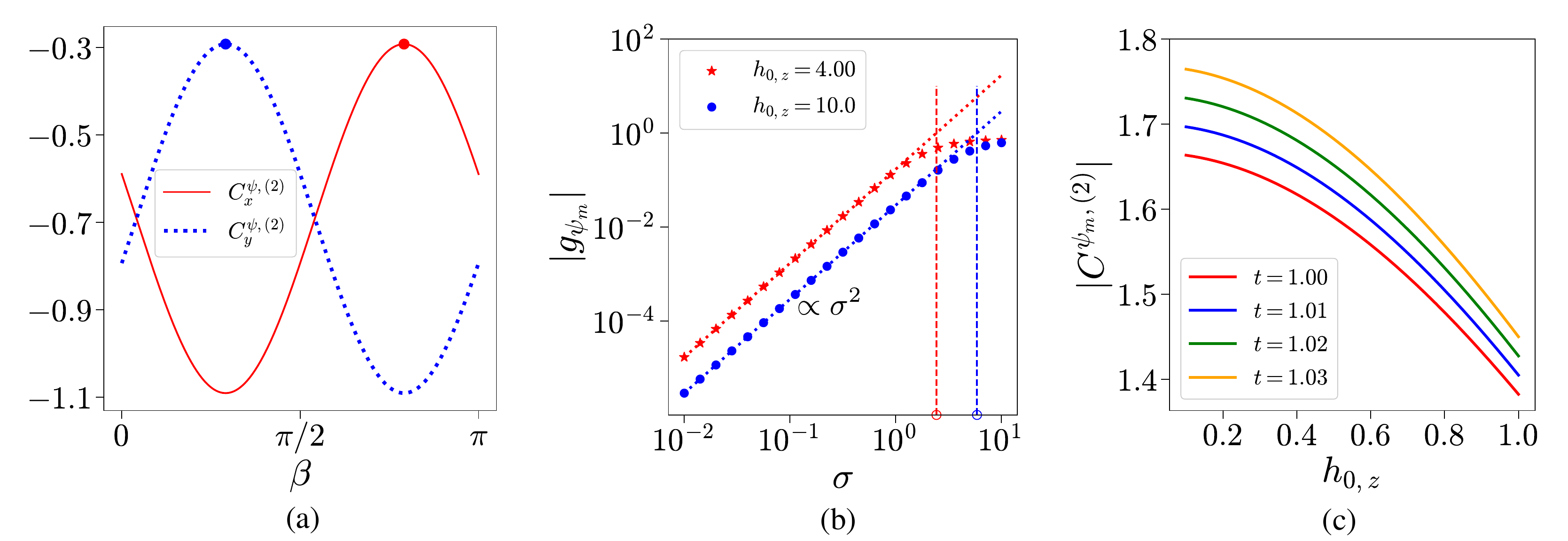}
    \caption{\textbf{Single-qubit probe under disordered magnetic field.} (a) Variations of $C_x^{\psi_m,(2)}$ (solid line) and $C_y^{\psi_m,(2)}$ (dashed line) as functions of $\beta$ with $\alpha=\pi/2$, $h_{0,z}=1$, $t=1$. (b) Log-log plot of \(|g_{\psi_m}|\) versus \(\sigma_x=\sigma_y=\sigma\) for different \(h_{0,z}\). Points denote numerically evaluated values from Eq.~(\ref{eq:f_psi}) using disorder averaging over \(10^6\) realizations of \(H\) (Eq.~(\ref{eq:single_qubit_Hamiltonian})) with normally distributed disorder. Vertical dashed lines indicate \(\sigma_{\psi_m}^{\max}=2.426\,(2.423)\) and \(5.866\,(5.858)\) for \(h_{0,z}=4(10)\), calculated from $C^{\psi_m,(2)}$. (c) Variation of $|C^{\psi_m,(2)}|$ with $h_{0,z}$ for different values of $t$ and $\sigma_x=\sigma_y$.}
    \label{fig:qubit}
\end{figure*}

\subsection{The disorder marker} 
\label{subsec:disorder_marker}

We define the \emph{disorder marker} (DM) as
\begin{eqnarray}\label{eq:disorder_marker}
g_\psi=f_\psi-1,
\end{eqnarray}
where
\begin{eqnarray}\label{eq:f_psi}
f_\psi&=&\overline{F_\psi}/F_{0,\psi},\nonumber\\
g_\psi&=&\sum_{r=2,3,\cdots}\sum_n q_n^{(r)}C_n^{\psi,(r)},
\end{eqnarray}
with
\begin{eqnarray}\label{eq:C_n^(r)}
C_n^{\psi,(r)}
=
4\tilde{G}_n^{\psi,(r)}/F_{0,\psi}.
\end{eqnarray}
Notably, \(C_n^{\psi,(r)}\) can be computed entirely from \(H_0\) (see Supplementary Materials for details) without performing any disorder average. When \(g_\psi=0\), or equivalently \(\overline{F_\psi}=F_{0,\psi}\), disorder does not affect the measurement precision, yielding a \emph{disorder-immune probe} (DIP), equivalent to a CP. In general, disorder adversely affects the probe, leading to
\(0\leq \overline{F_\psi}\leq F_{0,\psi}\), or equivalently,
\(0\leq f_\psi<1\), resulting in a \emph{disorder-sensitive probe} (DSP). In this regime, \(g_\psi\) (\(-1\leq g_\psi<0\)) quantifies the robustness of the probe for a given initial state \(\ket{\psi}\): smaller \(|g_\psi|\) implies greater robustness, while \(|g_\psi|\to 1\) indicates strong sensitivity to disorder. Conversely, when \(f_\psi>1\), or equivalently \(g_\psi>0\), disorder enhances the QFI, yielding a \emph{disorder-enhanced probe} (DEP).

If the disorder distribution is symmetric about its mean, i.e.,
\(P_n(\phi_{0,n}+\delta\phi_n)=P_n(\phi_{0,n}-\delta\phi_n)\ \forall n\), all odd central moments vanish, and neglecting terms of order \(\mathcal{O}((\delta\phi_n)^r)\) with even \(r\geq4\) yields 
\begin{eqnarray}
g_\psi\approx \sum_n q_n^{(2)} C_n^{\psi,(2)}
=\sum_n \sigma_n^2 C_n^{\psi,(2)},
\label{eq:quantum_marker}
\end{eqnarray}
where \(q_n^{(2)}=\sigma_n^2\) is the variance of \(P_n(\phi_n)\), with \(\sigma_n>0\) (\(\sigma_n=0\ \forall n\) recovers the CP). For uniform disorder strength, \(\sigma_n=\sigma\ \forall n\),
\begin{eqnarray}\label{eq:f_symmetric}
\ln |g_\psi|
\approx
2\ln \sigma+\ln |C^{\psi,(2)}|,
\end{eqnarray}
where \(C^{\psi,(2)}=\sum_n C_n^{\psi,(2)}\). Thus, for a DSP, the maximum disorder strength sustainable without significant degradation is
\begin{eqnarray}\label{eq:max_sigma}
\sigma_\psi^{\max}
=
1/\sqrt{|C^{\psi,(2)}|},
\end{eqnarray}
where the subscript emphasizes the dependence on the choice of \(\ket{\psi}\). Hence, \(|C^{\psi,(2)}|\) provides an intrinsic measure of resilience of the probe against disorder, which is calculable solely from the spectrum of the CP (see Appendix~\ref{app:C_calculation}). One can further optimize robustness by minimizing \(|g_\psi|\) over all initial probe states, i.e., \(|g_{\psi_r}|=\min_{\{\ket{\psi}\}} |g_\psi|\), 
where \(\ket{\psi_r}\) denotes the maximally resilient initial probe state. Note that for symmetric and identical disorder, the DEP condition \(g_\psi>0\) reduces to \(C^{\psi,(2)}>0\), while for symmetric disorder with unequal strengths, \(\sum_n \sigma_n^2 C_n^{\psi,(2)}>0\), which may arise from several possible sign combinations of \(C_n^{\psi,(2)}\). 

For disorder distributions asymmetric about their means,
\begin{eqnarray}
    g_\psi\approx \sum_n q_n^{(2)}C_n^{\psi,(2)}+\sum_n q_n^{(3)}C_n^{\psi,(3)},
\end{eqnarray}
where terms \(\mathcal{O}((\delta\phi_n)^r)\) with \(r\geq4\) are neglected. Using \(q_n^{(3)}=\gamma_n\sigma_n^3\), with \(\gamma_n\neq0\) the skewness of \(P_n(\phi_n)\), and assuming \(\sigma_n=\sigma\) and \(\gamma_n=\gamma\ \forall n\), one obtains 
\begin{eqnarray}
    g_\psi\approx \sigma^2 C^{\psi,(2)}[1+C_{32}^\psi\gamma\sigma],
\end{eqnarray}
where \(C^{\psi,(3)}=\sum_n C_n^{\psi,(3)}\) and \(C_{32}^\psi=C^{\psi,(3)}/C^{\psi,(2)}\). Thus, the effect of disorder asymmetry is negligible when \(C_{32}^\psi\gamma\sigma\ll1\), i.e., \(\gamma\sigma\ll1/C_{32}^\psi\), providing a criterion to assess the impact of asymmetric disorder on probe performance. However, we specifically focus on the symmetric disorder in the rest of the paper.  Note further that the Cram\'{e}r-Rao bound~\cite{Rao1945,Cramer1946,Helstrom1967,Helstrom1969,Holevo1982,Braunstein1994,Paris2009} provides a lower bound of the uncertainty in parameter estimation with $\nu$ independent trials, given by $\Delta \theta^{\mathrm{CR}}\geq 1/\sqrt{\nu F_{\psi}}$, which modifies to $\Delta\theta^{\mathrm{CR}}_0 \geq 1/\sqrt{\nu F_{0,\psi}}$ in the disorder-free situation, where $\nu$ denotes the number of independent trials. In the presence of a non-zero $g_{\psi}$, $\Delta\theta^{\mathrm{CR}}
=\Delta\theta^{\mathrm{CR}}_{0}/\sqrt{1+g_{\psi}}$. For a DIP, $\Delta\theta^{\mathrm{CR}}
=\Delta\theta^{\mathrm{CR}}_{0}$, while for a DSP (DEP),  $\Delta\theta^{\mathrm{CR}}>\Delta\theta^{\mathrm{CR}}_{0}$ ($\Delta\theta^{\mathrm{CR}}<\Delta\theta^{\mathrm{CR}}_{0}$).

\section{Operational perspective}
\label{sec:operational_perspective}

In practice, an experiment with the designed probe, initialized in the pure state $\ket{\psi}$ and undergoing a time evolution followed by a projective measurement  $\{M_x=\ket{m_x}\bra{m_x}\}$ with $\sum_{x}M_x=I$ produces a random outcome $x$ drawn from the distribution $p(x|\theta)$~\cite{Helstrom1967,Braunstein1994, Yu2022,Yu2021}. For $\nu$ independent trials,  the uncertainty in estimating $\theta$ from these outcomes is bounded as $\Delta\theta\geq 1/\sqrt{\nu F_{c,\psi}}$, where $F_{c,\psi}$ is the classical Fisher information (CFI)~\cite{Fisher1922}, defined as
\begin{eqnarray}
    F_{c,\psi}&=&\sum_x\frac{\left[\partial_\theta p(x|\theta)\right]^2}{p(x|\theta)},
\end{eqnarray}
assuming that each $p(x|\theta)>0$. Using 
\begin{eqnarray}
    P_x\equiv p(x|\theta)&=&\langle \psi_t|M_x|\psi_t\rangle=\langle\psi|M_{x,t}|\psi\rangle,
\end{eqnarray}
where $M_{x,t}=U_t^\dagger M_x U_t$ (see Secs.~\ref{sec:intro} and \ref{sec:methodology}), and the definition of the QFIG, one obtains
\begin{eqnarray}
\label{eq:CFI_main}
    F_{c,\psi}&=&\sum_x P_x^{-1}\left[\text{i}\bra{\psi}[G,M_{x,t}]\ket{\psi}\right]^2.
\end{eqnarray}
In contrast, the QFI, $F_\psi$, depends on $\ket{\psi_t}$ alone, and not on the measurement performed. The bridge between the CFI and the QFI is the Hermitian operator $L_\theta$, referred to as the symmetric logarithmic derivative (SLD)~\cite{Carollo2018, Helstrom1967, Helstrom1969, Braunstein1994}, satisfying 
\begin{eqnarray}
    \partial_\theta\rho_\theta=\frac{1}{2}\left\{\rho_\theta,L_\theta\right\},
\end{eqnarray}
for a parameter-encoded state $\rho_\theta$. For the pure state $\rho_\theta=\ket{\psi_t}\bra{\psi_t}$ considered in this paper (see Secs.~\ref{sec:intro} and \ref{sec:methodology}), 
\begin{eqnarray}
    L_\theta= 2\partial_{\theta}[\ket{\psi_t}\bra{\psi_t}]=-2\text{i}\left[U_tGU_t^\dagger,\rho_\theta\right]. 
\end{eqnarray}
The CFI and the QFI satisfy~\cite{Braunstein1994}
\begin{eqnarray}
    F_{c,\psi}\left[\{M_x\}\right]\leq F_\psi=\max_{\{M_x\}}F_{c,\psi}\left[\{M_x\}\right], 
    \label{eq:CFI-QFI-hierarchy}
\end{eqnarray}
for every pure state $\ket{\psi_t}$ and every measurement $\{M_x\}$.  The eigenbasis of $L_\theta$ (see Appendix~\ref{app:sld_basis}) corresponds to one (but not the only one) measurement for which $F_{c,\psi}=F_\psi$.

Similar to the QFI, the  CFI, quenched averaged over independent disorder distributions with zero mean and upon keeping terms up to the second standardized moment of the disorder distribution can be written as (see Appendix~\ref{app:cfi_expansion} for details)  
\begin{eqnarray}
    \overline{F_{c,\psi}} &\approx& F_{c,\psi}^{(0)}\left[1+\sum_{n}\sigma_n^2C_{c,n}^{\psi,(2)}\right],
\end{eqnarray}
with $C_{c,n}^{\psi,(2)}=F_{c,\psi,n}^{(2)}/F_{c,\psi}^{(0)}$, where  $F_{c,\psi}^{(0)}\equiv F_{c,0,\psi}$, the zero'th order term in the expansion, is the CFI for the clean probe for the same measurement (we have modified the notation to keep it uniform with the QFI).  For brevity, we refrain from including the expression for $F_{c,\psi,n}^{(2)}$ in the main text; it can be found in Appendix~\ref{app:cfi_expansion}. This allows one to define the \emph{classical} disorder marker as 
\begin{eqnarray}
    g_{c,\psi}&=&\frac{\overline{F_{c,\psi}}}{F_{c,0,\psi}}-1,
\end{eqnarray}
which, for symmetric disorder distributions, become (see Eq.~(\ref{eq:quantum_marker}) for comparison) 
\begin{eqnarray}
    g_{c,\psi}&\approx&\sum_{n}\sigma_n^2C_{c,n}^{\psi,(2)},
    \label{eq:classical_marker}
\end{eqnarray}
For equal disorder strength $\sigma_n=\sigma$ $\forall n$, we define  $C_c^{\psi,(2)}=\sum_{n}C_{c,n}^{\psi,(2)}$, leading to  $g_{c,\psi}=\sigma^2C_c^{\psi,(2)}$. For a DSP ($C_{c}^{\psi,(2)}<0$),   the disorder strength at which the leading correction becomes as large as the clean
value is (see Eq.~(\ref{eq:max_sigma}) for comparison)
\begin{eqnarray}
    \sigma^{\max}_c=1/\sqrt{|C_c^{\psi(2)}|},
    \label{eq:max_sigma_classical}
\end{eqnarray}
beyond which the second-order truncation is no longer meaningful.  Calculation of $C_c^{\psi(2)}$, similar to $C^{\psi(2)}$, requires the spectrum of the clean probe alone (see Appendix~\ref{app:C_classical_calculation} for details).

The experimentally relevant situation is the one  where the same \emph{fixed} measurement, $\{M_x=\ket{m_x}\bra{m_x}\}$, is used for every disorder realization, since the experimentalist has no  access to information regarding the disorder realization. Here, ``fixed" refers to the laboratory (Schr\"{o}dinger picture) operator $M_x$, which is independent of $\{\delta\phi_n\}$, while the Heisenberg picture operator, $M_{x,t}$, depends on the disorder through $U_t$, which is captured in Appendix~\ref{app:cfi_expansion}. We further assume that the clean probabilities corresponding to this measurement are strictly positive (this corresponds to $P_x^{(0)}>0$ $\forall x$, where $P_x^{(0)}$ is the zero'th order term in the expansion of $P_x$ -- see Appendix~\ref{app:cfi_expansion}). Since $\ket{\psi_t}$ is pure for every disorder realization, Eq.~(\ref{eq:CFI-QFI-hierarchy}) implies $F_{c,\psi}[\{\delta\phi_n\}]\leq F_{\psi}[\{\delta\phi_n\}]$ for each disorder realization $\{\delta\phi_n\}$, and each fixed $M_x$, while  for the clean probe (i.e., $\delta\phi_n=0\forall n$), $F_{c,0,\psi}\leq F_{0,\psi}$ irrespective of the disorder distribution. We refer to a fixed measurement $M_x$ as a \emph{clean optimal measurement} (COM) if $F_{c,0,\psi}=F_{0,\psi}$.

For a specific disorder realization $\{\delta\phi_n\}$, let us define
\begin{eqnarray}
    D[\{\delta\phi_n\}]=F_\psi[\{\delta\phi_n\}]-F_{c,\psi}[\{\delta\phi_n\}],
    \label{eq:difference}
\end{eqnarray}
with $D[\{\delta\phi_n\}]\geq 0$ for all realizations $\{\delta\phi_n\}\in\mathbb{R}^N$, and $D[\mathbf{0}]=F_{0,\psi}-F_{c,0,\psi}=0$ (the clean probe) if and only if the fixed measurement is COM. We choose to work with a COM, for which $D[\{\delta\phi_n\}]$ attains its minimum at the origin. The condition $P_x^{(0)}>0$ ensures $P_x>0$, and hence the smoothness of $D[\{\delta\phi_n\}]$ in the neighborhood of $\mathbf{0}$. This allows one to probe the behavior of $D[\{\delta\phi_n\}]$ in the vicinity of the origin along any line through it.

For independent disorder distributions with zero mean, the quenched average of $F_\psi$ and $F_{c,\psi}$, up to second order, involve only $\tilde{G}^{\psi,(2)}_n$ and $F_{c,n}^{\psi,(2)}$, i.e., the diagonal coefficients, which are fixed numbers independent of which disorder realization is chosen. It is, therefore, sufficient to  choose a realization in which only $\delta\phi_n=\varepsilon\neq 0$ while $\delta\phi_{m}=0$ $\forall m\neq n$. Along this line,  
\begin{eqnarray}
    D(\varepsilon)=\lambda_1\varepsilon+\lambda_2\varepsilon^2+D^\prime(\varepsilon),
    \label{eq:D_epsilon}
\end{eqnarray}
with $\lambda_1=4\tilde{G}_n^\psi-F_{c,\psi,n}^{(1)}$, $\lambda_2=4\tilde{G}_n^{\psi,(2)}-F_{c,\psi,n}^{(2)}$, and  $D^\prime(\varepsilon)\sim\mathcal{O}(|\varepsilon|^3)$. Since $D(\varepsilon)\geq 0$ and $D(0)=0$, the origin is a minimum of $D(0)$, i.e., $D(\varepsilon)$ can neither have a non-zero slope, or a negative curvature there, leading to $\lambda_1=0$ and $\lambda_2\geq 0$ (see Appendix~\ref{app:proof_of_hierarchy} for an explicit proof), and subsequently, 
\begin{eqnarray}
    \label{eq:lambda_12}
    4\tilde{G}_n^\psi&=&F_{c,\psi,n}^{(1)},\;4\tilde{G}_n^{\psi,(2)}\geq F_{c,\psi,n}^{(2)}.  
\end{eqnarray}
Dividing by $F_{c,0,\psi}=F_{0,\psi}>0$, one obtains 
\begin{eqnarray}
    C_{c,n}^{\psi,(2)}\leq C_{n}^{\psi,(2)},
    \label{eq:C_hierarchy}
\end{eqnarray}
and subsequently (since $\sigma^2_n\geq 0$),
\begin{eqnarray}
    g_{c,\psi}\approx \sum_{n}\sigma^2_n C_{c,n}^{\psi,(2)}\leq \sum_{n}\sigma^2_{n}C_{n}^{\psi,(2)}\approx g_{\psi}. 
    \label{eq:g_hierarchy}
\end{eqnarray}
For equal disorder strengths, $C_c^{\psi,(2)}\leq C^{\psi,(2)}$. In the case of DSPs, $C_{c}^{\psi,(2)}\leq C^{\psi,(2)}< 0$, leading to $|C_{c}^{\psi,(2)}|\geq |C^{\psi,(2)}|$, and hence $\sigma^{\max}_{c,\psi}\leq  \sigma^{\max}_\psi$. This implies that the laboratory-realized probe is never more robust than what is predicted by the QFI, as CFI can never be less affected by disorder than the QFI. Therefore, a probe that is DSP w.r.t. the QFI remains DSP w.r.t. the CFI  also.   In contrast, a DIP or a DEP w.r.t. the QFI ($C^{\psi,(2)}\geq 0$) may turn into a DSP w.r.t. the CFI ($C_c^{\psi,(2)}<0$), but not the other way around.  

Note that to establish the above hierarchy, COM is essential. For instance, with $F_{c,0,\psi}<F_{0,\psi}$ (i.e., when the fixed measurement is not COM), $D(\mathbf{0})>0$, and $D(\varepsilon)=D(\mathbf{0})+\lambda_1\varepsilon+\lambda_2\varepsilon^2+D^\prime(\varepsilon)$ is positive irrespective of the values of $\lambda_{1,2}$. Further, $C_{c,n}^{\psi,(2)}$ and $C_{n}^{\psi,(2)}$ are normalized by different values as $F_{c,0,\psi}\neq F_{0,\psi}$.

\section{Robustness of a single-qubit probe} 
\label{sec:single_qubit}

We first consider a single-qubit probe governed by
\begin{eqnarray}
H=h_{0,z}Z+\vec{\delta h}\cdot\vec{P},
\label{eq:single_qubit_Hamiltonian}
\end{eqnarray}
where \(\vec{P}=(X,Y,Z)\) are the Pauli matrices, \(h_{0,z}\) is the intended magnetic field along the \(z\)-direction to be estimated, and \(\vec{\delta h}=(\delta h_x,\delta h_y,\delta h_z)\) represents disorder arising from stray magnetic fields, leading to \(H_0=h_{0,z}Z\) with \(\theta\equiv h_{0,z}\), and \(V=\vec{\delta h}\cdot\vec{P}\). Each \(h_n\) (\(n=x,y,z\)) follows a disorder distribution with mean \(h_{0,n}\) and standard deviation \(\sigma_n\), determined by the probe design and fabrication. In the present setup, \(h_{0,n}=0\) for \(n=x,y\).

\subsection{The QFI-maximized probe}
\label{subsec:qfi_disorder}

We consider the QFI-maximized single-qubit probe in the CP limit, motivating the study of \(g_{\psi_m}\). Using the \(Z\)-eigenbasis $Z\ket{0}=+\ket{0},\;Z\ket{1}=-\ket{1}$, \(\ket{\psi_m}=(\ket{0}+\mathrm{e}^{\mathrm{i}\beta}\ket{1})/\sqrt{2}\), lying on the \(xy\)-plane of the Bloch sphere.

Explicit calculation leads to $C_z^{\psi_m,(2)}=0$, and, for \(n=x,y\),
\begin{eqnarray}\label{eq:single_qubit_C_xy}
    C_n^{\psi_m,(2)}
&=&-\frac{1}{4h_{0,z}^4t^2}\Bigg[2\left(
\cos 2h_{0,z}t+2h_{0,z}^2t^2-1
\right)\nonumber\\
&&+\bigg\{2h_{0,z}t\left(\delta_{n,y}\sin\beta+\delta_{n,x}\cos\beta\right)\nonumber\\
&&+\delta_{n,y}\cos(2h_{0,z}t+\beta)
-\delta_{n,x}\sin(2h_{0,z}t+\beta)\nonumber\\
&&-\delta_{n,y}\cos\beta+\delta_{n,x}\sin\beta\bigg\}^2\Bigg],
\end{eqnarray}
where \(\delta_{n,n'}\) is the Kronecker delta. The probe, therefore, becomes a DIP when disorder is present only along the \(z\)-field. Moreover, \(C_x^{\psi_m,(2)}\) and \(C_y^{\psi_m,(2)}\) both are \(\pi\)-periodic in \(\beta\), and shifted by \(\pi/2\) on $\beta$-axis relative to each other (see Fig.~\ref{fig:qubit}(a)).

\emph{Does the choice of $\ket{\psi_m}$ as the initial probe state ensure maximal robustness of the probe against disorder}? Note that for \(\sigma_x=\sigma_y\), optimizing \(g_{\psi_m}\) is equivalent to optimizing
\(C^{\psi_m,(2)}=C_x^{\psi_m,(2)}+C_y^{\psi_m,(2)}\), which is independent of \(\beta\). However, when \(\sigma_x\neq\sigma_y\), this symmetry is broken, and the optimal choice of \(\beta\) is determined by optimizing \(g_{\psi_m}\) w.r.t. $\beta$, depending on the hierarchy between \(\sigma_x\) and \(\sigma_y\). For instance, for a DSP (\(0<|g_{\psi_m}|\leq1\)) with $\sigma_x>\sigma_y$, maximal resilience is achieved by choosing $\beta$ in $\ket{\psi_m}$ as 
\begin{eqnarray}
    \beta^m=\tan^{-1}\left[(2^{-1}\sin 2h_{0,z}t-h_{0,z}t)/\sin^2 h_{0,z}t\right],
\end{eqnarray}
minimizing \(|g_{\psi_m}|\).

For $\sigma_x=\sigma_y=\sigma$, in Fig.~\ref{fig:qubit}(b) of the main text, we plot the variation of $\left|g_{\psi_m}\right|=\left|1-f_{\psi_m}\right|$ as a function of $\sigma$, where $\overline{F_{\psi_m}}$ in $f_{\psi_m}$ is computed via performing a quenched average of $F_{\psi_m}=4\Delta^2_{\psi_m}G$ over $10^6$ realizations of the disordered Hamiltonian (Eq.~(\ref{eq:single_qubit_Hamiltonian})). The data clearly demonstrates a $\sim\sigma^2$ variation of $\left|g_{\psi_m}\right|$ (see Eq.~(\ref{eq:f_symmetric})) when $\sigma$ is small. The maximum values, $\sigma^{\max}_{\psi_m}$ (see Eq.~(\ref{eq:max_sigma}) of the main text), of $\sigma$, as estimated via calculating $\left|C^{\psi_m,(2)}\right|$ from the CP using the prescription given in Appendix.~\ref{app:C_calculation}, are also shown, which are in agreement with the calculated values of $\left|C^{\psi_m,(2)}\right|$ by fitting the disorder-averaged data to Eq.~(\ref{eq:f_symmetric}) of the main text. A deviation from $\sim\sigma^2$ variation is observed as $\sigma\rightarrow\sigma^{\max}_{\psi_m}$, as $\left|g_{\psi_m}\right|$ starts saturating to $1$. Eq.~(\ref{eq:single_qubit_C_xy}) implies that  $g_{\psi_m}=\sigma_x^2C^{\psi_m,(2)}_x+\sigma_y^2C^{\psi_m,(2)}_y\leq0$ irrespective of the choice of $\sigma_x,\sigma_y,h_{0,z},t$ and $\beta$. Therefore the single-qubit probe defined by Eq.~(\ref{eq:single_qubit_Hamiltonian}) is always disorder-sensitive. The  maximum sustainable disorder strength $\sigma^{\max}_{\psi_m}$ with $h_{0,z}$ can be extracted from the variation of $|C^{\psi_m,(2)}|=|C^{\psi_m,(2)}_x+C^{\psi_m,(2)}_y|$ with $h_{0,z}$, which exhibits an increasingly faster decay with increasing $h_{0,z}$, as demonstrated in Fig.~\ref{fig:qubit}(c).

\subsection{QFI-maximized probe with COM}
\label{subsec:cfi_disorder}

We now consider the probe prepared in the state $\ket{\psi_m}$, followed by a time evolution generated by $H$, and subsequently a measurement given by $\{M_{\pm}\}$. An arbitrary single-qubit measurement can be written as
\begin{eqnarray}
    M_{\pm}&=&\frac{1}{2}\left[I\pm \hat{\mathbf{n}}.\vec{P}\right], 
\end{eqnarray}
where
\begin{eqnarray}
    \hat{\mathbf{n}}&=&(\sin\vartheta\cos\varphi,\ \sin\vartheta\sin\varphi,\ \cos\vartheta),
\end{eqnarray}
with two real parameters: the polar angle $\vartheta\in[0,\pi]$, and the azimuth $\varphi\in[0,2\pi]$. The measurement is \emph{fixed} when the same $(\vartheta,\varphi)$ is used for all disorder realization. 

\begin{figure*}
    \centering
    \includegraphics[width=0.8\linewidth]{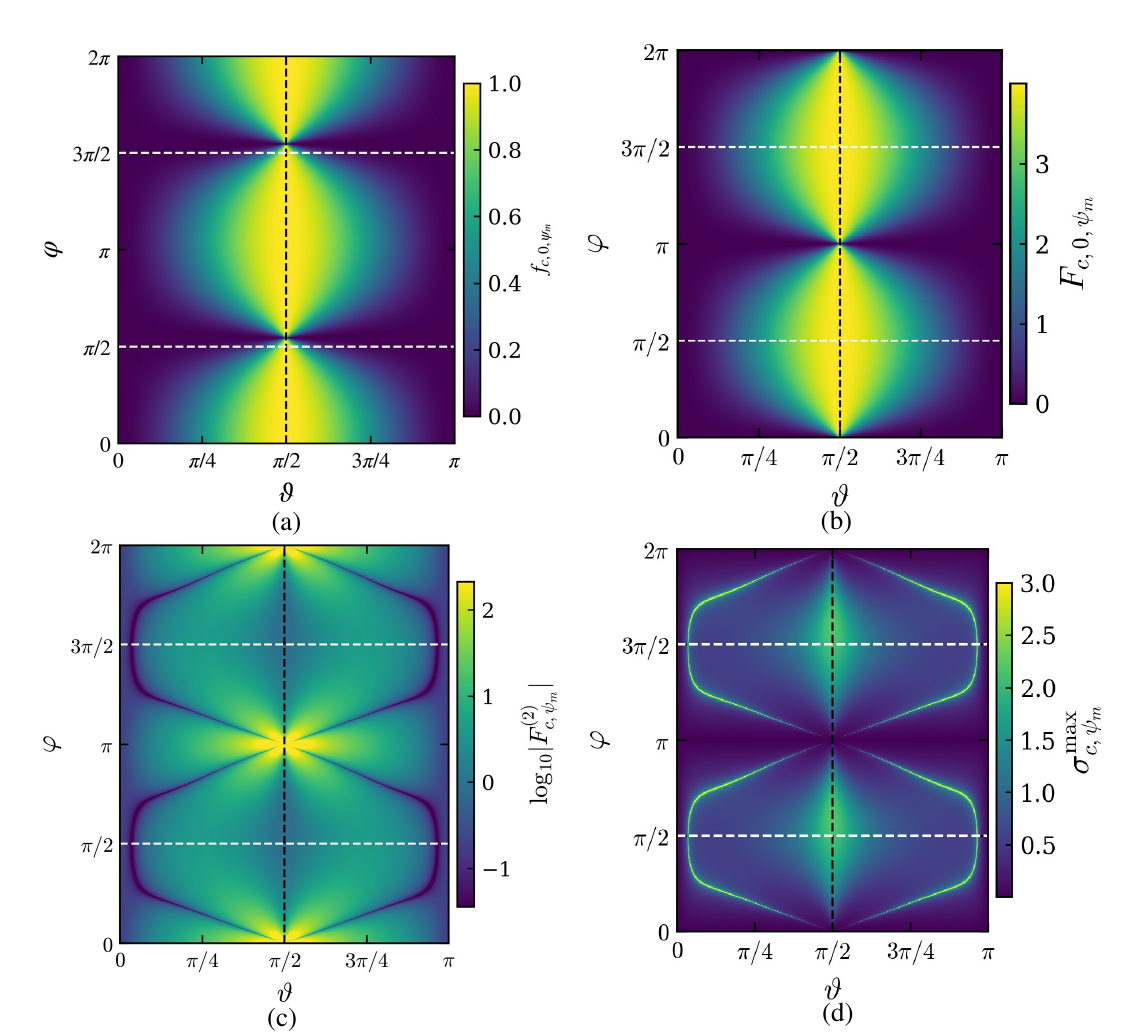}
    \caption{\textbf{Measurement landscape.}  Variations of (a) $f_{c,0,\psi_m}$, (b) $F_{c,0,\psi_m}$, (c) $\log_{10}\big|F^{(2)}_{c,\psi_m}\big|$, and (d) $\sigma^{\max}_{c,\psi_m}$ as functions of $\vartheta$ and $\varphi$.  The clean SLD measurements correspond to points $(\pi/2,\pi/2)$ and $(\pi/2,3\pi/2)$. We have chosen $\sigma=0.5$, $\beta=0$, $h_{0,z}=4$, and $t=1$ for (a)-(d).}
    \label{fig:qubit_measurement}
\end{figure*}

In the disorder-free scenario, the time-evolved QFI-optimized state, up to a global phase $\exp (-\text{i}h_{0,z}t)$, reads 
\begin{eqnarray}
    \ket{\psi_t}=\frac{1}{\sqrt{2}}\left(\ket{0}+\text{e}^{\text{i}\nu_t}\ket{1}\right),
\end{eqnarray}
corresponding to the Bloch vector $\vec{\mathbf{r}}=(\cos\nu_t,\sin\nu_t,0)$, where $\nu_t=\beta+2h_{0,z}t$ is a time-dependent local phase. Upon measurement, the probabilities of outcomes are $p_{\pm}=(1\pm\hat{\mathbf{n}}.\vec{\mathbf{r}})/2$, using which one obtains 
\begin{eqnarray}
    F_{c,0,\psi_m}&=&\frac{\left(\hat{\mathbf{n}}.\partial_{h_{0,z}}\vec{\mathbf{r}}\right)^2}{1-\left(\hat{\mathbf{n}}.\vec{\mathbf{r}}\right)^2}=\frac{4t^2\sin^2\vartheta\sin^2\xi_t}{1-\sin^2\vartheta\cos^2\xi_t},
\end{eqnarray}
where $\xi_t=\varphi-\nu_t$ is the time-dependent angle between the measurement axis and $\vec{\mathbf{r}}$. On the other hand, $F_{0,\psi_m}=4\Delta^2_{\psi_m}G^{(0)}$ with $G^{(0)}=\int_{0}^tds U_{0,s}^\dagger Z U_{0,s}=tZ$. Since $\langle Z\rangle=0$ and $\langle Z^2\rangle=1$ in $\ket{\psi_m}$, $F_{0,\psi_m}=4t^2$. Therefore, 
\begin{eqnarray}
    f_{c,0,\psi_m}=\frac{F_{c,0,\psi_m}}{F_{0,\psi_m}}=\frac{\sin^2\vartheta\sin^2\xi_t}{1-\sin^2\vartheta\cos^2\xi_t}\leq 1.
    \label{eq:com_test}
\end{eqnarray}
Equality in Eq.~(\ref{eq:com_test}), which requires $\vartheta=\pi/2$, i.e., measurement along the equatorial plane (see Fig.~\ref{fig:qubit_measurement}(a) for the variation of $f_{c,0,\psi_m}$ as functions of $\vartheta$ and $\varphi$). Further, $\xi_t\neq 0,\pi$ in order for keeping $p_\pm>0$. Therefore, COMs are characterized by $\vartheta=\pi/2$, $\xi_t\in(0,\pi)$. For clean SLD, $\varphi=\pi/2$ (see Appendix~\ref{app:sld_basis}), implying $\xi_t=(\pi/2-\beta) -2h_{0,z}t$.

It is important to note that the construction of the probe Hamiltonian as well as the choice of the measurement to estimate the parameter depend on both the maximum sustainable disorder strength quantifying the robustness, and the value of  the Fisher information -- quantum, or classical -- bounding the precision of parameter estimation. To demonstrate this, we plot, in Fig.~\ref{fig:qubit_measurement}(b), the variation of $F_{c,0,\psi_m}$ as a function of $(\vartheta,\varphi)$, which determines the (classical) precession bound for the designed probe. Further, in Fig.~\ref{fig:qubit_measurement}(c) and (d), we respectively plot $\log_{10}\big|F^{(2)}_{c,\psi_m}\big|$ and $\sigma^{\max}_{c,\psi_m}$, as functions of  $(\vartheta,\varphi)$, demonstrating existence of measurements for which the maximum sustainable disorder strength can be large due to a sharp decrease in $\big|F^{(2)}_{c,\psi_m}\big|$, and yet $F_{c,0,\psi_m}$is substantially low, indicating a significant decrease in the measurement precision.

\section{Disordered Kitaev model as a probe}
\label{sec:kitaev_model}

Next, we consider a disordered multi-qubit probe to demonstrate the utility of the disorder marker. Motivated by recent interest in many-body quantum metrology~\cite{Paris2016,Hauke2016,Frerot2018,Ilias2022,Ding2022,Angelo2022,DiCandia2023,Montenegro2025}, we choose the one-dimensional quadratic Kitaev model~\cite{Kitaev2001} with nearest-neighbor (NN) interactions,
\begin{eqnarray}
\label{eq:disordered_Kitaev}
H&=&\sum_{i=1}^{N-1}
-\tau_i(c_i^\dagger c_{i+1}+c_{i+1}^\dagger c_i)
+\eta_i(c_i c_{i+1}+ c_{i+1}^\dagger c_i^\dagger)\nonumber\\
&&-\mu\sum_{i=1}^{N}
\left(c_i^\dagger c_i-\frac{1}{2}\right),
\end{eqnarray}
where \(\tau_i\) and \(\eta_i\) are the NN hopping and pairing amplitudes, respectively, \(\mu\) is the uniform disorder-free chemical potential, and \(c_i^\dagger\) (\(c_i\)) are fermionic creation (annihilation) operators satisfying \(\{c_i,c_j^\dagger\}=\delta_{ij}\). We take \(\mu\equiv\theta\) as the parameter to be estimated, while \(\tau_i=\tau_{0,i}+\delta\tau_i\) and \(\eta_i=\eta_{0,i}+\delta\eta_i\) (\(\equiv\phi\)) undergo site-dependent fluctuations, introducing spatial inhomogeneity. Also, we use open boundary conditions for this paper, although the qualitative results remain unchanged for periodic boundary conditions. Note that the model maps to several paradigmatic 1D spin systems with NN interactions~\cite{Pfeuty1970,Pfeuty1971,Stinchcombe1973,Stinchcombe1973a,Stinchcombe1973b,Barouch1970,Barouch1971,Barouch1971a,Lieb1961} via the Jordan-Wigner transformation~\cite{Wigner1928,Fradkin1989,Batista2001,Sachdev2011}.

\begin{figure*}
    \centering
    \includegraphics[width=0.7\linewidth]{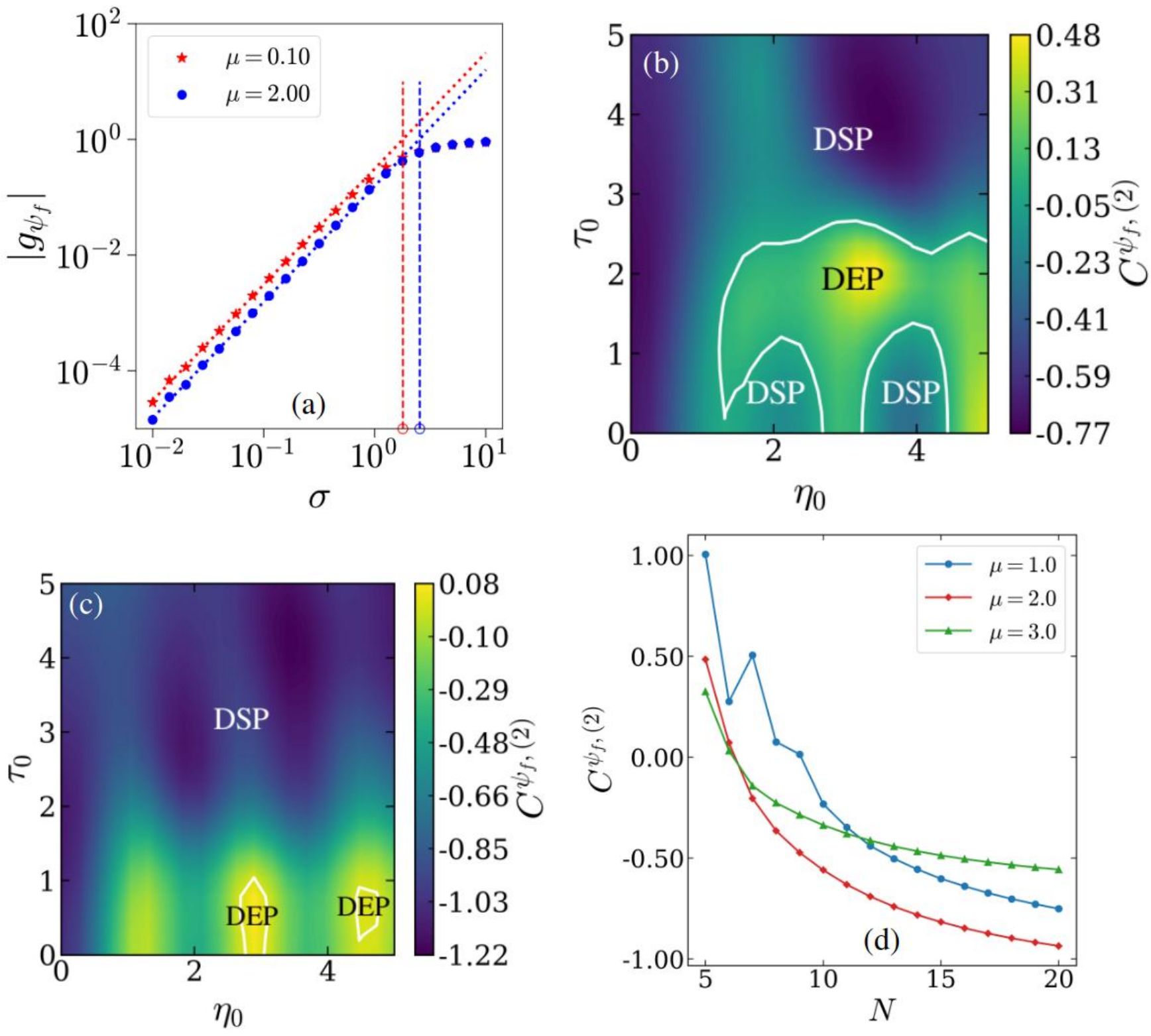}
    \caption{\textbf{Disordered Kitaev probe} (a) Log-log plot of $|g_{\psi_f}|$ with $\sigma$ for $N=6$ for different $\mu$, keeping $\tau_0=\eta_0=-1$, $t=1$. The numerical data (denoted by points) are obtained via quenched averaging over $10^7$ realizations of $H$ (Eq.~(\ref{eq:disordered_Kitaev})), for identical Gaussian disorder distributions. The vertical dashed lines $\sigma^{\max}_{\psi_f}=1.792(2.544)$ corresponding to $\mu=0.01(2.0)$ are obtained by directly computing $|C^{\psi_f,(2)}|$, which agree with the $\sigma^{\max}_{\psi_f}$ ($1.793$ and $2.522$, respectively) obtained by fitting Eq.~(\ref{eq:f_symmetric}) to the numerical data. (b)-(c)  Landscapes of $C^{\psi_f,(2)}$ on the $(\eta_0,\tau_0)$ plane keeping (b) $N=5$ and (c) $N=20$, with the contours correspond to $C^{\psi_f,(2)}=0$. (d) Variations of $C^{\psi_f,(2)}$ against $N$ for $\mu=1,2$, and $3$, where the corresponding values of $(\tau_0,\eta_0)$ are fixed at $(1.84,5)$, $(2.11,3.42)$, and $(2.63,3.16)$ respectively, where the maximum $C^{\psi_f,(2)}$  corresponding to the specific values of $\mu$ at $N=5$ are obtained.}
    \label{fig:kitaev}
\end{figure*}

The disordered Hamiltonian in Eq.~(\ref{eq:disordered_Kitaev}) is quadratic in fermionic operators and can therefore be written in the Bogoliubov-de Gennes (BdG) form~\cite{deGennes1999,Sadhukhan2016},
\begin{eqnarray}
H=\sum_{i,j=1}^{N} c_i^\dagger A_{ij} c_j
+\frac{1}{2}\sum_{i,j=1}^{N}
\left(
c_i^\dagger B_{ij} c_j^\dagger+\text{h.c.}
\right),
\end{eqnarray}
where \(A\) (\(B\)) is a symmetric (antisymmetric) \(N\times N\) matrix describing hopping (pairing), with
\begin{eqnarray}
A_{ij}&=&-[\tau_i\delta_{i+1,j}+\tau_j\delta_{i,j+1}+\mu\delta_{i,j}],\nonumber\\
B_{ij}
&=&
-[\eta_i\delta_{i+1,j}-\eta_j\delta_{i,j+1}],
\end{eqnarray}
for \(i,j<N\), together with
\(A_{1N}=A_{N1}=0\), \(A_{NN}=-\mu\), and
\(B_{1N}=B_{N1}=B_{NN}=0\). Introducing the Nambu spinor~\cite{Nambu1960}
\begin{eqnarray}\label{eq:nambu_spinor}
\mathbf{c}
=
\begin{pmatrix}
c_1 & \cdots & c_N &
c_1^\dagger & \cdots & c_N^\dagger
\end{pmatrix}^T,
\end{eqnarray}
the Hamiltonian and the QFIG become
\begin{eqnarray}\label{eq:H_in_terms_of_M}
H&=&\frac{1}{2}\mathbf{c}^\dagger M\mathbf{c},\nonumber\\ 
G&=&\frac{1}{2}\mathbf{c}^\dagger J\mathbf{c},
\end{eqnarray}
where the BdG matrix
\begin{eqnarray}\label{eq:bdg_matrices}
M=
\begin{pmatrix}
A & B \\
- B & -A
\end{pmatrix},
\end{eqnarray} 
and 
\begin{eqnarray}\label{eq:J}
    J=
\int_0^t ds\,
\mathrm{e}^{\mathrm{i}Ms}
M_\mu
\mathrm{e}^{-\mathrm{i}Ms},
\end{eqnarray}
with 
\begin{eqnarray}
M_\mu=
\begin{pmatrix}
-I_N & 0 \\
0 & I_N
\end{pmatrix},
\end{eqnarray}
and \(I_N\) is the \(N\times N\) identity matrix.

Motivated by its Heisenberg scaling in the TFIM for all interaction and field strengths~\cite{Skotiniotis2015} and its experimental accessibility~\cite{Sackett2000,Leibfried2005,Pan2000,Neeley2010,Wei2005,Riedel2010,Gross2010,Bernien2017}, we choose the \(N\)-qubit Greenberger-Horne-Zeilinger (GHZ) state~\cite{GHZ1989},
\begin{eqnarray}
    \ket{\psi_c}
=\big(
\ket{0}^{\otimes N}
+\ket{1}^{\otimes N}
\big)/\sqrt{2},
\end{eqnarray} 
as the initial CP state, which, in the fermionic basis~\cite{Wigner1928,Lieb1961}, become
\begin{eqnarray}
    \ket{\psi_f}
=\left(
\ket{\mathbf{v}}
+\ket{\mathbf{f}}
\right)/\sqrt{2},
\end{eqnarray}
where \(c_i\ket{\mathbf{v}}=0\ \forall i\), and
\(\ket{\mathbf{f}}=\prod_{i=1}^{N}c_i^\dagger\ket{\mathbf{v}}\).
Since \(G\) (\(G^2\)) is quadratic (quartic) in fermionic operators,
\(\langle \mathbf{v}|G|\mathbf{f}\rangle=0\) for \(N>2\) and
\(\langle \mathbf{v}|G^2|\mathbf{f}\rangle=0\) for \(N>4\), due to orthogonality between different particle-number sectors. Hence, for \(N>4\),
\begin{eqnarray}
    \langle G\rangle_{\psi_f}
&=&(\langle G\rangle_{\mathbf{v}}
+\langle G\rangle_{\mathbf{f}})/2,\nonumber\\
\langle G^2\rangle_{\psi_f}
&=&(\langle G^2\rangle_{\mathbf{v}}
+\langle G^2\rangle_{\mathbf{f}})/2,
\end{eqnarray}
leading to
\begin{eqnarray}
    \Delta_{\psi_f}^2G
=
\big[(\Delta_{\mathbf{v}}^2G
+\Delta_{\mathbf{f}}^2G)/2
+
\left(
\langle G\rangle_{\mathbf{v}}
-\langle G\rangle_{\mathbf{f}}
\right)^2/4\big].
\end{eqnarray}

To obtain an explicit operator form of $G$, the time-integration in Eq.~(\ref{eq:J}) needs to be evaluated, which is convenient in eigenbasis of the full Hamiltonian via spectral decomposition.  
Introducing the quasi-particle operators $\mathbf{d}$ through a Bogoliubov transformation~\cite{Bogoliubov1958,Lieb1961,Altland2010}, given by $\mathbf{d}=\mathcal{B}^{-1}\mathbf{c}$, $M$ is diagonalized as $\mathcal{B}^{-1} M \mathcal{B} = \Lambda$, with $\Lambda=\text{diag}(\lambda_1,\ldots,\lambda_{2N})$. 
The operator $G$ can be expressed in this quasi-particle basis as
\begin{equation}
G=\frac{1}{2}\,\mathbf{d}^\dagger K \mathbf{d},
\end{equation}
with
\begin{equation}
K_{\alpha\beta}=
\frac{e^{i(\lambda_\alpha-\lambda_\beta)t}-1}{i(\lambda_\alpha-\lambda_\beta)}
\left[\mathcal{B}^{-1}M_\mu\mathcal{B}\right]_{\alpha\beta}.
\end{equation}
In the degenerate limit $\lambda_\alpha\to\lambda_\beta$, this reduces to
\begin{equation}
K_{\alpha\beta}=t\left[\mathcal{B}^{-1}M_\mu\mathcal{B}\right]_{\alpha\beta}.
\end{equation}
Inverting the Bogoliubov transformation as $\mathbf{c}=\mathcal{B}\mathbf{d}$ yields
\begin{equation}
G=\frac{1}{2}\,\mathbf{c}^\dagger\left(\mathcal{B}K\mathcal{B}^{-1}\right)\mathbf{c},
\end{equation}
leading to the effective matrix representation of the operator $J$ in the fermionic basis  as
\begin{equation}
J=\mathcal{B}K\mathcal{B}^{-1}.
\end{equation}

While the GHZ state $\ket{\psi_f}$ is non-Gaussian~\cite{Blaizot1986,Peschel2003,Bravyi2005}, both $\ket{\mathbf{v}}$ and $\ket{\mathbf{f}}$ are Gaussian states, allowing all expectation values to be  determined by the two-point correlation matrix~\cite{Blaizot1986,Peschel2003,Peschel2009}, defined in the Nambu-space~\cite{Blaizot1986} as $R_{ij}=\langle \mathbf{c}_j \mathbf{c}_i^\dagger \rangle_\psi$, which, corresponding to $\ket{\mathbf{v}}$ and $\ket{\mathbf{f}}$, becomes 
\begin{equation}
R_{\mathbf{v}}=
\begin{pmatrix}
I & 0\\
0 & 0
\end{pmatrix}, \quad
R_{\mathbf{f}}=
\begin{pmatrix}
0 & 0\\
0 & I
\end{pmatrix}.
\end{equation}
The expectation value of $G$ w.r.t. the GHZ state, in terms of the correlation matrix, reads 
\begin{equation}
\langle G \rangle_\psi = \frac{1}{2}\text{Tr}\left[J(I-R)\right],
\end{equation}
while 
\begin{equation}
\langle G^2 \rangle_\psi = \frac{1}{4}\sum_{ijkl} J_{ij}J_{kl}\langle \mathbf{c}_i^\dagger \mathbf{c}_j \mathbf{c}_k^\dagger \mathbf{c}_l \rangle_\psi.
\end{equation}
Using Wick’s theorem~\cite{Blaizot1986,Peschel2003,Peschel2009},
\begin{eqnarray}
\langle \mathbf{c}_i^\dagger \mathbf{c}_j \mathbf{c}_k^\dagger \mathbf{c}_l \rangle_\psi
&=&
\langle \mathbf{c}_i^\dagger \mathbf{c}_j \rangle_\psi \langle \mathbf{c}_k^\dagger \mathbf{c}_l \rangle_\psi
-\langle \mathbf{c}_i^\dagger \mathbf{c}_k^\dagger \rangle_\psi \langle \mathbf{c}_j \mathbf{c}_l \rangle_\psi\nonumber\\&&
+\langle \mathbf{c}_i^\dagger \mathbf{c}_l \rangle_\psi \langle \mathbf{c}_j \mathbf{c}_k^\dagger \rangle_\psi,
\end{eqnarray}
leading to 
\begin{eqnarray}
\langle G^2 \rangle_\psi
&=&
\frac{1}{4}\bigg[\left(\text{Tr}[J(I-R)]\right)^2
+\text{Tr}[JRJ]
\nonumber\\&&-\text{Tr}[JRJR]
-\sum_{ijkl} J_{ij}J_{kl}
\langle \mathbf{c}_i^\dagger \mathbf{c}_k^\dagger \rangle_\psi
\langle \mathbf{c}_j \mathbf{c}_l \rangle_\psi\bigg],
\end{eqnarray}
and subsequently,
\begin{eqnarray}
\Delta^2_\psi G
&=&
\frac{1}{4}\bigg[\mathrm{Tr}[JRJ(I-R)]
\nonumber\\&&-\sum_{ijkl} J_{ij}J_{kl}
\langle \mathbf{c}_i^\dagger \mathbf{c}_k^\dagger \rangle_\psi
\langle \mathbf{c}_j \mathbf{c}_l \rangle_\psi\bigg].
\end{eqnarray}
This allows one to calculate the expectation value and the variance of $G$ with respect to $\ket{\mathbf{v}}$ and $\ket{\mathbf{f}}$ in the Nambu-space as 
\begin{eqnarray}
\langle G \rangle_{\mathbf{v}(\mathbf{f})} &=& \frac{1}{2}\text{Tr}\left[J(I-R_{\mathbf{v}(\mathbf{f})})\right],\\
\Delta^2_{\mathbf{v}(\mathbf{f})} G
&=&
\frac{1}{4}\bigg[\mathrm{Tr}[JR_{\mathbf{v}(\mathbf{f})}J(I-R_{\mathbf{v}(\mathbf{f})})]\nonumber\\&&
-\sum_{ijkl} J_{ij}J_{kl}
\langle \mathbf{c}_i^\dagger \mathbf{c}_k^\dagger \rangle_{\mathbf{v}(\mathbf{f})}
\langle \mathbf{c}_j \mathbf{c}_l \rangle_{\mathbf{v}(\mathbf{f})}\bigg].
\end{eqnarray} 
Using these, we numerically evaluate \(|g_{\psi_f}|\) in the disordered Kitaev model for \(\sigma_\tau=\sigma_\eta=\sigma\), assuming \(\tau_{0,i}=\tau_0\) and \(\eta_{0,i}=\eta_0\ \forall i\), and for different \(\mu\), \(\tau_0\), and \(\eta_0\). Figure~\ref{fig:kitaev}(a) shows \(|g_{\psi_f}|\sim\sigma^2\) for \(\sigma\ll\sigma_{\psi_f}^{\max}\), obtained via disorder averaging over \(10^7\) realizations of Eq.~(\ref{eq:disordered_Kitaev}).

In order to directly evaluate $C^{\psi_f,(2)}$ using $H_0,V$ and $H_\theta$ (see Appendix.~\ref{app:C_calculation}), one requires expectation values of the product of two general quadratic fermion operators, say, $A$ and $B$, given by 
\begin{eqnarray}
   \braket{AB}_\mathbf{v} &=&\frac{1}{4}\bigg\{\mathrm{Tr}[A(I-R_\mathbf{v})]\mathrm{Tr}[B(I-R_\mathbf{v})]\nonumber\\&&+\mathrm{Tr}[B(I-R_\mathbf{v})AR_\mathbf{v}]-\sum_{il}A_{i,l-N}B_{i-N,l} \bigg\},\nonumber\\
   \braket{AB}_\mathbf{f} &=&\frac{1}{4}\bigg\{\mathrm{Tr}[A(I-R_\mathbf{f})]\mathrm{Tr}[B(I-R_\mathbf{f})]\nonumber\\&&+\mathrm{Tr}[B(I-R_\mathbf{f})AR_\mathbf{f}]-\sum_{kj}A_{k-N,j}B_{k,j-N} \bigg\},\nonumber\\
\end{eqnarray}
calculated in $\ket{\mathbf{v}}$ and $\ket{\mathbf{f}}$ in the Nambu-space. At fixed \(t\) and \(N\), numerical evaluation of \(C^{\psi_f,(2)}\) over the parameter space \((\mu,\tau_0,\eta_0)\) reveals finite regions with \(C^{\psi_f,(2)}>0\), corresponding to a DEP, while the surfaces \(C^{\psi_f,(2)}=0\) separating the DEP from the DSP \((C^{\psi_f,(2)}<0)\) form \emph{planes of immunity} hosting the DIP. Figures~\ref{fig:kitaev}(b), (c) show projections of the parameter space for \(N=5,20\) at fixed \(\mu=2\) and \(t=1\). The DEP region shrinks with increasing \(N\), indicating the suppression of disorder-enhanced QFI for larger system sizes. This is in agreement with the overall decay of $C^{\psi_f,(2)}$ against $N$ for fixed points $(\mu,\tau_0,\eta_0)$ in the parameter space (see Fig.~\ref{fig:kitaev}(d)). When the disorder strengths are identical for all parameters, \(\sigma_{\psi_f}^{\max}\) can be directly estimated from Eq.~(\ref{eq:max_sigma}) as a function of \((\mu,\tau_0,\eta_0)\).

\section{Conclusion and outlook}
\label{sec:conclude}

In this paper, we investigated the impact of glassy disorder in system parameters on the quantum Fisher information for parameter estimation, where the treatment is completely independent of the platform with which the probe is fabricated. Via a Dyson expansion of the quantum Fisher information of the disordered probe in terms of the different orders of standardized central moments of the relevant disorder distributions, we have defined a disorder marker quantifying the sole effect of the disorder. Using the disorder marker, we have quantitatively categorized the different effect of disorder on a quantum probe, namely, the disorder-immune, disorder-sensitive, and disorder-enhanced probes. We have proved that in the cases of symmetric and identical disorders with the same disorder strength affecting all parameters of the probe-Hamiltonian, the absolute value of the disorder marker for a disorder-sensitive probe varies quadratically with disorder strength. We have further shown that the maximum value of the admissible disorder strength that the probe can withstand competes with an intrinsic robustness scale that can be computed from the Hamiltonian of the clean (disorder-free) probe without performing any disorder-averaging. We have extended our calculation to scenarios where a measurement, fixed irrespective of the disorder realization, is performed for estimating the desired parameter, and have shown that a specific hierarchy between the maximum sustainable disorder strength of the probe can be obtained as long as the chosen measurement leads to identical classical and quantum Fisher information in the disorder-free case. We have demonstrated the power of our results in the case of a single-qubit probe under disordered magnetic field, and a multi-qubit probe described by a disordered Kitaev model with nearest-neighbor interactions. 

These results enable the systematic design of robust many-body probes whose metrological advantages, such as Heisenberg scaling~\cite{Leibfried2004} or quantum-criticality-enhanced sensing~\cite{Zanardi2008}, can survive sufficiently weak disorder. The resulting disorder thresholds also motivate exploring many-body localization~\cite{Basko2006,Nandkishore2015} and glassy phases~\cite{Sherrington1975,Edwards1975,Derrida1981,Cugliandolo2003} as metrological resources, and allows benchmarking disorder-induced performance degradation in platform-specific quantum sensors. Beyond metrology, it also provides a route toward disorder-aware quantum control, where optimal control protocols~\cite{Doria2011} can be designed using disorder thresholds to preserve coherence and sensitivity.

\acknowledgements 

A.K.P and V.K. acknowledge the support from the Anusandhan National Research Foundation (ANRF) of the Department of Science and Technology (DST), India, through the Core Research Grant (CRG) (File No. CRG/2023/001217, Sanction Date 16 May 2024). H.K.J. acknowledges the Prime Minister Research Fellowship Program, Government of India, for the financial support. The Authors thank Keshav Das Agarwal and Aditi Sen(De) for pointing out errors in references. 

\newpage

\onecolumngrid

\appendix

\section{Expansion of quantum Fisher information generator}
\label{app:dyson_expansion}

Invoking Dyson-series expansion~\cite{Dyson1949,SakuraiQM} of $\mathcal{U}_s$ as
\begin{equation}\label{Dyson-series}
\mathcal{U}_s=\sum_{k=0,1,2,\cdots}\mathcal{U}_s^{(k)},
\end{equation}
with
\begin{eqnarray}
    \mathcal{U}_s^{(0)}&=&I, \label{U_0}\nonumber \\ 
    \mathcal{U}_s^{(1)}&=&-\text{i}\int_0^s ds_1\mathcal{V}_{s_1},\label{U_1}\nonumber \\ 
    \mathcal{U}_s^{(2)}&=&-\int_0^{s}ds_1\int_0^{s_1}ds_2 \mathcal{V}_{s_1}\mathcal{V}_{s_2}, \label{U_2}\\
    &\vdots& \nonumber
\end{eqnarray}
where  $\mathcal{V}_s=U_{0,s}^\dagger VU_{0,s}$, a similar expansion of $G$, given by
\begin{equation}
G=\sum_{k=0,1,2,\cdots}G^{(k)},
\end{equation}
can be obtained. Defining $\mathcal{H}_{\theta,s}=U_{0,s}^\dagger\partial_{\theta}H U_{0,s}$,  the $k$th order contributions to $G$ is 
\begin{eqnarray}
G^{(k)}=\sum_{\underset{a+b=k}{a,b}}\int_0^t ds \mathcal{U}_s^{(a)\dagger}\mathcal{H}_{\theta,s}\mathcal{U}_s^{(b)}, 
\end{eqnarray}
the explicit forms of which are given up to $k=3$ below.
\begin{enumerate}
    \item $k=0$:
    \begin{eqnarray}
    G^{(0)}=\int_0^t ds\mathcal{H}_{\theta,s}. 
    \end{eqnarray}

    \item $k=1$: Defining $\mathcal{H}_{n,s}=U_{0,s}^\dagger H_n U_{0,s}$, 
    \begin{eqnarray}
    G^{(1)}&=&\text{i}\sum_n\delta \phi_n \int_0^t ds\int_0^sds_1\left[\mathcal{H}_{n,s_1},\mathcal{H}_{\theta,s}\right].
    \end{eqnarray}

    \item $k=2$:  
    \begin{eqnarray}    
    G^{(2)} &=& \sum_{n_1,n_2}\delta\phi_{n_1}\delta\phi_{n_2}\int_0^t  ds \left(\int_0^s ds_1\mathcal{H}_{n_1,s_1}  \right)\mathcal{H}_{\theta,s}\left(\int_0^s ds_1 \mathcal{H}_{n_2,s_1}  \right)\nonumber\\&& 
    -\sum_{n_1,n_2}\delta\phi_{n_1}\delta\phi_{n_2}\int_0^t ds \int_0^s ds_1 \int_0^{s_1} ds_2   \left(
    \mathcal{H}_{n_1,s_2}\mathcal{H}_{n_2,s_1}\mathcal{H}_{\theta,s} + \mathcal{H}_{\theta,s}\mathcal{H}_{n_2,s_1} \mathcal{H}_{n_1,s_2}
    \right).
    \end{eqnarray}   

    \item $k=3$:
    \begin{eqnarray}
        G_{n_1,n_2,n_3}^{(3)}&=& -\text{i}\sum_{n_1,n_2,n_3}\delta\phi_{n_1}\delta\phi_{n_2}\delta\phi_{n_3}\int_0^t  ds \left(\int_0^s ds_1 \mathcal{H}_{n_1,s_1}\right) \mathcal{H}_{\theta,s}\left(\int_0^s ds_1 \int_0^{s_1} ds_2 \mathcal{H}_{n_2,s_1}\mathcal{H}_{n_3,s_2}  \right)\nonumber\\
        &&+\text{i}\sum_{n_1,n_2,n_3}\delta\phi_{n_1}\delta\phi_{n_2}\delta\phi_{n_3}\int_0^t  ds \left(\int_0^s ds_1 \int_0^{s_1} ds_2 \mathcal{H}_{n_2,s_1}\mathcal{H}_{n_3,s_2}  \right) \mathcal{H}_{\theta,s}\left(\int_0^s ds_1 \mathcal{H}_{n_1,s_1}\right)\nonumber\\
        &&+\text{i}\sum_{n_1,n_2,n_3}\delta\phi_{n_1}\delta\phi_{n_2}\delta\phi_{n_3}\int_0^t  ds \int_0^s ds_1 \int_0^{s_1} ds_2 \int_0^{s_2} ds_3 \mathcal{H}_{\theta,s}\mathcal{H}_{n_1,s_1}\mathcal{H}_{n_2,s_2}\mathcal{H}_{n_3,s_3} \nonumber\\
        &&-\text{i}\sum_{n_1,n_2,n_3}\delta\phi_{n_1}\delta\phi_{n_2}\delta\phi_{n_3}\int_0^t  ds \int_0^s ds_1 \int_0^{s_1} ds_2 \int_0^{s_2} ds_3 \mathcal{H}_{n_3,s_3}\mathcal{H}_{n_2,s_2}\mathcal{H}_{n_1,s_1}\mathcal{H}_{\theta,s}.
    \end{eqnarray}
\end{enumerate}
This leads to 
\begin{eqnarray}
    \langle G \rangle&=&\langle G^{(0)}\rangle+\sum_n \delta \phi_n \langle G_n^{(1)}\rangle+\sum_{n_1,n_2}\delta \phi_{n_1} \delta \phi_{n_2} \langle G_{n_1n_2}^{(2)}\rangle
    +\sum_{n_1,n_2,n_3}\delta \phi_{n_1} \delta \phi_{n_2} \delta \phi_{n_3} \langle G_{n_1n_2n_3}^{(3)}\rangle+\cdots
\end{eqnarray}
and 
\begin{eqnarray}
    \langle G^2 \rangle &=&\langle [G^{(0)}]^2\rangle+\sum_n \delta \phi_n \left[\langle G^{(0)} G_n^{(1)}\rangle+\langle G_n^{(1)} G^{(0)} \rangle \right]+\sum_{n_1,n_2} \delta \phi_{n_1} \delta\phi_{n_2} \bigg[\langle G_{n_1}^{(1)} G_{n_2}^{(1)}\rangle+\langle G^{(0)} G_{n_1n_2}^{(2)}\rangle+\langle G_{n_1n_2}^{(2)}G^{(0)} \rangle \bigg]\nonumber\\
    &&+\sum_{n_1,n_2,n_3} \delta \phi_{n_1} \delta\phi_{n_2}\delta\phi_{n_3} \bigg[\langle G_{n_1}^{(1)} G_{n_2n_3}^{(2)}\rangle+\langle G_{n_1n_2}^{(2)} G_{n_3}^{(1)}\rangle+\langle G^{(0)} G_{n_1n_2n_3}^{(3)}\rangle+\langle G_{n_1n_2n_3}^{(3)}G^{(0)} \rangle \bigg],
\end{eqnarray}
resulting in the variance of quantum Fisher information generator as
\begin{eqnarray}
\Delta^2 G &=& \Delta^2G^{(0)} + \sum_n\delta\phi_n \tilde{G}_n + \sum_{n_1,n_2}\delta\phi_{n_1}\delta\phi_{n_2} \tilde{G}_{n_1n_2}+\sum_{n_1,n_2,n_3}\delta\phi_{n_1}\delta\phi_{n_2} \delta\phi_{n_3} \tilde{G}_{n_1n_2n_3}+\cdots, 
\end{eqnarray}
with
\begin{eqnarray}
\label{eq:G_tilde}
    \tilde{G}_n&=&\langle G^{(0)} G_n^{(1)}\rangle+\langle G_n^{(1)} G^{(0)} \rangle-2\langle G^{(0)}\rangle\langle G_n^{(1)}\rangle,\nonumber\\
    \tilde{G}_{n_1n_2}&=&\langle G_{n_1}^{(1)} G_{n_2}^{(1)}\rangle+\langle G^{(0)} G_{n_1n_2}^{(2)}\rangle+\langle G_{n_1n_2}^{(2)}G^{(0)} \rangle-2\langle G^{(0)}\rangle\langle G_{n_1n_2}^{(2)}\rangle-\langle G_{n_1}^{(1)}\rangle\langle G_{n_2}^{(1)}\rangle,\nonumber\\
    \tilde{G}_{n_1n_2n_3}&=&\langle G^{(0)} G_{n_1n_2n_3}^{(3)}\rangle+\langle G_{n_1n_2n_3}^{(3)}G^{(0)} \rangle
    +\langle G_{n_1}^{(1)}G_{n_2n_3}^{(2)} \rangle+\langle G_{n_2n_3}^{(2)}G_{n_1}^{(1)} \rangle - 2\langle G^{(0)}\rangle\langle G_{n_1n_2n_3}^{(3)}\rangle-2\langle G_{n_1}^{(1)}\rangle\langle G_{n_2n_3}^{(2)}\rangle,
\end{eqnarray}
where we have included terms up to third order in fluctuations.

\section{Calculation of \texorpdfstring{$C_n^{\psi,(2)}$}{C2} from the spectrum of clean-probe Hamiltonian}
\label{app:C_calculation}

We work in the energy eigenbasis of the disorder-free Hamiltonian $H_0$ satisfying $H_0\ket{E_k}=E_k\ket{E_k}$, such that $H_0 = S D S^\dagger$, with 
where $D = \text{diag}(E_1, E_2, \cdots)$, and $S$ is the unitary matrix diagonalizing $H_0$. Defining
\begin{equation}
\widetilde{\partial_{\theta} H} = S^\dagger (\partial_{\theta} H) S,
\end{equation}
one obtains
\begin{eqnarray}
    \mathcal{H}_{\theta,s}
    &=& S e^{\text{i}Ds} \widetilde{\partial_{\theta} H} e^{-\text{i}Ds} S^\dagger.
\end{eqnarray}
Thus, in the eigenbasis of $H_0$,
\begin{equation}
  \left[\partial_{\theta} H^\prime\right]_{ij}
  = \left[\widetilde{\partial_{\theta} H}\right]_{ij} e^{\text{i}\Delta E_{ij}s},
\end{equation}
where $\Delta E_{ij}=E_i - E_j$, and  $\left[.\right]_{ij}$ is the $i,j$th matrix element. Similarly, one defines  $\widetilde{H}_n = S^\dagger H_n S$ and obtain
\begin{equation}
  \left[H_n^\prime\right]_{ij} = \left[\widetilde{H}_n\right]_{ij} e^{\text{i}\Delta E_{ij}s}.
\end{equation}
Defining 
\begin{eqnarray}
T_{ij}&=&\left\{\begin{array}{ll}
        -\text{i}\Delta E_{ij}^{-1}\left(\text{e}^{\text{i}\Delta E_{ij}t}-1\right), & \text{if }  i\neq j,\\[6pt]
        t, & \text{if } i = j,
    \end{array}\right.
    \label{eq:Tij}\\
S_{ijk}&=&\left\{\begin{array}{ll}
        \Delta E_{ik}^{-1}\left(T_{ij}-T_{kj}\right), & \text{if }  i\neq k,\\
    \Delta E_{ij}^{-2}\left[t\Delta E_{ij}e^{\text{i}\Delta E_{ij}t}+\text{i}(\text{e}^{\text{i}\Delta E_{ij}t}-1)\right], & \text{if } i=k \text{ and }i\neq j \\
        \text{i}t^2/2,&  \text{if }  i=k \text{ and }i=j
    \end{array}\right.\\
R_{ijkl}&=&\left\{\begin{array}{ll}
        \Delta E_{ik}^{-1}\left(S_{ijl}-S_{kjl}\right), & \text{if }  i\neq k\\
        
        \Delta E_{ij}^{-2} \Delta E_{il}^{-2} \Delta E_{jl}^{-1}\bigg[\text{i} \text{e}^{\text{i} \Delta E_{lj} t} \Delta E_{ij}^2-\text{i} \Delta E_{il}^2 & \\
        \quad\quad +\text{e}^{\text{i} \Delta E_{ij} t} \Delta E_{jl}\left\{\text{i}(\Delta E_{ij}+\Delta E_{il})+\Delta E_{ij}\Delta E_{il}t\right\}\bigg], & \text{if } i=k \text{ and }i\neq j,i\neq l,j\neq l\\
        
        \Delta E_{li}^{-3}\left[2\text{i} +\Delta E_{li}t + \text{e}^{\text{i} \Delta E_{il} t} \left(\Delta E_{li}t -2\text{i}\right)\right], & \text{if } i=k \text{ and }i\neq j,i\neq l,j= l\\
        
        2^{-1}\text{i}\Delta E_{ij}^{-3}\left[2 + \text{e}^{\text{i} \Delta E_{ij} t} \left(-2 +2\text{i}t\Delta E_{ij} +\Delta E_{ij}^2t^2\right)\right],& \text{if } i=k \text{ and }i\neq j,i= l\\
        
        2^{-1}\text{i}\Delta E_{il}^{-1} t^2-\Delta E_{il}^{-3}\left[\text{i}\left(1-\text{e}^{-\text{i} \Delta E_{il} t}\right) + \Delta E_{il}t\right], & \text{if } i=k \text{ and }i= j,i\neq l\\
        - t^3/6, & \text{if } i=k \text{ and }i= j,i= l
    \end{array}\right.\\
\overline{R}_{ijkl}&=&\left\{\begin{array}{ll}
        \Delta E_{ik}^{-1}\left(S_{lij}-S_{lkj}\right), & \text{if }  i\neq k\\
        
        \Delta E_{ij}^{-2} \Delta E_{il}^{-2} \Delta E_{jl}^{-1}\bigg[\text{i} (\Delta E_{ij}+\Delta E_{il})\Delta E_{jl} & \\
        \quad\quad+ \text{e}^{\text{i} \Delta E_{ji}t} \Delta E_{il}^2 \big(-\text{i}+\Delta E_{ij}t\big) +\text{e}^{\text{i} \Delta E_{li} t} \Delta E_{ij}^2\big(\text{i}- \Delta E_{il}t\big)\bigg], & \text{if } i=k \text{ and }i\neq j,i\neq l,j\neq l\\
        
        \text{i}\Delta E_{il}^{-3}\left[2+\text{e}^{\text{i} \Delta E_{li} t} \left(-2 + 2\text{i}t\Delta E_{li}+\Delta E_{li}^2t^2\right)\right], & \text{if } i=k \text{ and }i\neq j,i\neq l,j= l\\
        
        2^{-1}\Delta E_{ij}^{-3}\left[2\text{e}^{-\text{i} \Delta E_{ij} t} \left( \text{i} -  \Delta E_{ij}t\right)- \text{i} \left(2 + \Delta E_{ij}^2 t^2\right)\right],& \text{if } i=k \text{ and }i\neq j,i= l\\
        
        2^{-1} \Delta E_{il}^{-3}\left[2\text{e}^{-\text{i} \Delta E_{il} t} \left( \text{i} - \Delta E_{il}t\right)- \text{i} \left(2 + \Delta E_{il}^2 t^2\right)\right], & \text{if } i=k \text{ and }i= j,i\neq l\\
        t^3/3, & \text{if } i=k \text{ and }i= j,i= l
    \end{array}\right.    
\end{eqnarray}
with $T_{ji}=T_{ij}^{*}$ and $S_{ijk}=S_{kji}$, one obtains 
\begin{eqnarray}
   \left[G^{(0)}\right]_{ij} &=& \left[\widetilde{\partial_{\theta}H}\right]_{ij}T_{ij},
   \label{G0}\\
   \left[G_n^{(1)}\right]_{ij} &=& \sum_k\bigg\{\left[\widetilde{H}_n\right]_{ik}\left[\widetilde{\partial_{\theta}H}\right]_{kj}S_{ijk}+\left[\widetilde{\partial_{\theta}H}\right]_{ik}\left[\widetilde{H}_n\right]_{kj}S^*_{jik}\bigg\}, 
   \label{G1}\\
   \left[G_n^{(2)}\right]_{ij} &=& \sum_{kl}\bigg\{\left[\widetilde{H}_n\right]_{ik}\left[\widetilde{H}_n\right]_{kl} \left[\widetilde{\partial_{\theta}H}\right]_{lj}R_{ijkl}+\left[\widetilde{H}_n\right]_{ik} \left[\widetilde{\partial_{\theta}H}\right]_{kl}\left[\widetilde{H}_n\right]_{lj}\overline{R}_{ijkl}^{*}\nonumber\\&&+\left[\widetilde{\partial_{\theta}H}\right]_{ik}\left[\widetilde{H}_n\right]_{kl}\left[\widetilde{H}_n\right]_{lj}R_{lijk}^{*}\bigg\}.
   \label{G2}
\end{eqnarray}
Substituting these in Eq.~(\ref{eq:G_tilde}) yields $\tilde{G}_{nn}$, which results in $C_n^{\psi,(2)}$, and subsequently $C^{\psi,(2)}$ (see the main text). Note that estimating the effect of asymmetry in a disorder distribution requires the evaluation of $G_{n}^{(3)}$, given by
\begin{eqnarray}
G_{n}^{(3)}&=&\sum_{klm}\Bigg[\left(\Delta E_{ik}\Delta E_{mj}\Delta E_{lj}\right)^{-1}[\widetilde{H}_n]_{ik}[\widetilde{\partial_{\theta}H}]_{kl}[\widetilde{H}_n]_{lm}[\widetilde{H}_n]_{mj}(T_{ij}-T_{il}-T_{kj}+T_{kl})\nonumber\\
    &&-\left(\Delta E_{ik}\Delta E_{mj}\Delta E_{lm}\right)^{-1}[\widetilde{H}_n]_{ik}[\widetilde{\partial_{\theta}H}]_{kl}[\widetilde{H}_n]_{lm}[\widetilde{H}_n]_{mj}(T_{im}-T_{il}-T_{km}+T_{kl})\nonumber\\
    &&-\left(\Delta E_{kl}\Delta E_{mj} \Delta E_{il}\right)^{-1}[\widetilde{H}_n]_{ik}[\widetilde{H}_n]_{kl}[\widetilde{\partial_{\theta}H}]_{lm}[\widetilde{H}_n]_{mj}(T_{ij}-T_{lj}-T_{im}+T_{lm})\nonumber\\
     &&-\left(\Delta E_{mj}\Delta E_{lj} \Delta E_{kj}\right)^{-1}[\widetilde{\partial_{\theta}H}]_{ik}[\widetilde{H}_n]_{kl}[\widetilde{H}_n]_{lm}[\widetilde{H}_n]_{mj}(T_{ij}-T_{lj}-T_{im}+T_{lm})\nonumber\\    
    &&+\left(\Delta E_{kl}\Delta E_{mj} \Delta E_{ik}\right)^{-1}[\widetilde{H}_n]_{ik}[\widetilde{H}_n]_{kl}[\widetilde{\partial_{\theta}H}]_{lm}[\widetilde{H}_n]_{mj}\bigg\{\text{i}(\Delta E_{ik}+\Delta E_{lj})^{-1}(1-\text{e}^{\text{i}(\Delta E_{ik}+\Delta E_{lj})t})\nonumber\\&&-\text{i}(\Delta E_{ik}+\Delta E_{lm})[1-\text{e}^{\text{i}(\Delta E_{ik}+\Delta E_{lm})t}]+T_{lm}-T_{lj}\bigg\}\Bigg],
\end{eqnarray}
which can be further simplified and evaluated following the same prescription.

\section{Calculation of the clean SLD basis}
\label{app:sld_basis}

The operator $L_{\theta}$ is defined as~\cite{Carollo2018, Helstrom1967, Helstrom1969, Braunstein1994}
\begin{eqnarray}
    \frac{\partial\rho_{\theta}}{\partial\theta}=\frac{1}{2}\left(L_{\theta}\rho_{\theta}+\rho_{\theta}L_{\theta}\right),
\end{eqnarray}
which, upon writing $\rho_{\theta}=\sum_i p_i\ketbra{\lambda_i}{\lambda_i}$ in spectral-decomposed form, leads to~\cite{Braunstein1994,Paris2009}
\begin{eqnarray}
    L_{\theta}&=&\sum_{j,k}\frac{2\bra{\lambda_j}\partial_{\theta}\rho_{\theta}\ket{\lambda_k}}{p_k+p_j}\ketbra{\lambda_j}{\lambda_k}.
\end{eqnarray}
In this paper, we consider $\rho_{\theta}=\ketbra{\psi_{\theta}}{\psi_{\theta}}$ to be a pure state, for which the SLD can be expressed in terms of the QFIG as $L_{\theta}=-2\text{i}[\tilde{G}, \rho_{\theta}]$, where $\tilde{G}=UGU^\dagger$.  To access the measurements $\{M_{x,t}\}$, note that it corresponds to the eigenbasis of 
$L_{\mathrm{eff}}=U^\dagger L_{\theta}U=-2\text{i}[G,\rho_{0}]$, 
where $\rho_{(0)}$ is the chosen initial probe state. Expanding the SLD
in powers of the disorder variables
as
\begin{equation}
    L_{\mathrm{eff}}
    =
    L^{(0)}
    +
    \sum_n\delta\phi_n L_n^{(1)}
    +
    \sum_{n}(\delta\phi_n)^2 L_{n}^{(2)}
    +\cdots,
\end{equation}
with 
\begin{eqnarray}
    L^{(k)}
    &=&
    -2i\left[G^{(k)},\rho_0\right],k=0,1,2,\cdots
\end{eqnarray}
identify $L^{(0)}$ as the clean SLD basis for a given probe Hamiltonian with initial probe state $\rho_0$. The clean probe Hamiltonian satisfies $G=G^{(0)}$ and $L_{\mathrm{eff}}=L^{(0)}$, with the higher-order contributions vanishing, making the clean SLD a clean optimal basis.

\section{Expansion of the classical Fisher information}
\label{app:cfi_expansion}

We write Eq.~(\ref{eq:CFI_main}) as 
\begin{eqnarray}
\label{CFI_app}
    F_{c,\psi}&=&\sum_x \frac{A_x^2}{P_x},
\end{eqnarray}
with $A_x=\left[\text{i}\bra{\psi}[G,M_{x,t}]\ket{\psi}\right]$ and $P_x=\bra{\psi}M_{x,t}\ket{\psi}$. Similar to $G$ (see Eqs.~(\ref{eq:G_Dyson})-(\ref{eq:H_ns})), we expand the measurement in Dyson series as 
\begin{eqnarray}     M_{x,t}&=&\sum_{k=0}^\infty M_{x,t}^{(k)},
\end{eqnarray}
where, up to second order,  
\begin{eqnarray} M_{x,t}^{(0)}&=&\mathcal{M}_{x,t}=U_{0,t}^\dagger M_x U_{0,t},\nonumber\\
M_{x,t}^{(1)}&=&\sum_n \delta\phi_n M_{x,t,n}^{(1)},\nonumber\\
M_{x,t}^{(2)}&=&\sum_{n_1,n_2}\delta\phi_{n_1}\delta\phi_{n_2}M_{x,t,n_1n_2}^{(2)},
\end{eqnarray}
with
\begin{eqnarray}
M_{x,t,n}^{(1)}&=&\mathrm{i} \int_0^t ds_1
\left[\mathcal{H}_{n,s_1},\mathcal{M}_{x,t}\right],\nonumber\\
M_{x,t,n_1n_2}^{(2)}
&=&
\left(\int_0^t ds_1\, \mathcal{H}_{n_1,s_1}\right)
\mathcal{M}_{x,t}
\left(\int_0^t ds_1\, \mathcal{H}_{n_2,s_1}\right)\nonumber\\&&-\int_0^t ds_1\int_0^{s_1} ds_2
\big(
\mathcal{H}_{n_1,s_2}\mathcal{H}_{n_2,s_1}\mathcal{M}_{x,t}
+\mathcal{M}_{x,t}\mathcal{H}_{n_2,s_1}\mathcal{H}_{n_1,s_2}
\big).
\end{eqnarray}
Note that since $M_{x,t}$ is unitary conjugate of $M_x$, it is a positive operator, and $\sum_xM_{x,t}=I$ for every order: $\sum_{x}M_{x,t}^{(k)}=0$ for $k\geq 1$. Plugging in $F_c$, one obtains 
\begin{eqnarray}
    F_{c,\psi}=\sum_{x}\frac{\sum_{k=0}^\infty A_x^{(k)}}{\sum_{k=0}^{\infty}P_x^{(k)}}, 
\end{eqnarray}
where 
\begin{eqnarray}
A_x^{(1)}&=&\sum_n \delta \phi_n A_{x,n}^{(1)},\nonumber\\ 
A_x^{(2)}&=&\sum_{n_1,n_2}\delta \phi_{n_1}\delta \phi_{n_2} A_{x,n_1n_2}^{(2)}
\end{eqnarray}
with
\begin{eqnarray}
\label{eq:AX}
    A_x^{(0)}&=&\text{i}\bra{\psi}[G^{(0)},M_{x,t}^{(0)}]\ket{\psi},\nonumber\\
    A_{x,n}^{(1)}&=&\text{i}\Big[\bra{\psi}[G^{(0)},M_{x,t,n}^{(1)}]\ket{\psi}+\bra{\psi}[G_n^{(1)},M_{x,t}^{(0)}]\ket{\psi}\Big],\nonumber\\
    A_{x,n_1n_2}^{(2)}&=&\text{i}\Big[\bra{\psi}[G_{n_1}^{(1)},M_{x,t,n_2}^{(1)}]\ket{\psi}+\bra{\psi}[G^{(0)},M_{x,t,n_1n_2}^{(2)}]\ket{\psi}+\bra{\psi}[G_{n_1n_2}^{(2)},M_{x,t}^{(0)}]\ket{\psi}\Big],
\end{eqnarray}
and 
\begin{eqnarray}
P_x^{(1)}&=&\sum_n \delta \phi_n P_{x,n}^{(1)},\nonumber\\
P_x^{(2)}&=&\sum_{n_1,n_2}\delta\phi_{n_1}\delta\phi_{n_2}P_{x,n_1n_2}^{(2)},
\end{eqnarray}
with
\begin{eqnarray}
\label{eq:Px}
    P_x^{(0)}&=&\bra{\psi}M_{x,t}^{(0)}\ket{\psi},\nonumber\\
    P_{x,n}^{(1)}&=&\bra{\psi}M_{x,t,n}\ket{\psi},\nonumber\\
    P_{x,n_1n_2}^{(2)}&=&\bra{\psi}M_{x,t,n_1n_2}\ket{\psi}.
\end{eqnarray}

Writing $P_x=P_x^{(0)}\left[1+\epsilon_x\right]$ with $\epsilon_x=\sum_{k=1}^{\infty}P_x^{(k)}/P_x^{(0)}$, and identifying $P_x^{(0)}>0$ along with $\epsilon_x<1$ in the week disorder limit (it is, however, important to mention here that for $P_x^{(0)}=0$, $\epsilon_x$ is not small, and the expansion of $(1+\epsilon_x)^{-1}$ is not allowed. Such outcomes need to be treated separately), 
\begin{eqnarray}
    P_x^{-1}=\left(1-\epsilon_x+\epsilon_x^2-\epsilon_x^3+\cdots\right)/P_x^{(0)}.
\end{eqnarray} 
Retaining terms up to second order in fluctuations, 
\begin{eqnarray}
    \frac{1}{P_x}\approx\frac{1}{P_x^{(0)}}-\frac{P_x^{(1)}}{\left[P_x^{(0)}\right]^2}+\left[\frac{\left[P_x^{(1)}\right]^2}{\left[P_x^{(0)}\right]^3}-\frac{P_x^{(2)}}{\left[P_x^{(0)}\right]^2}\right]. 
    \label{eq:denominator}
\end{eqnarray}
Further, the numerator of $F_c$, up to second order in fluctuations, is
\begin{eqnarray}
    A_x^2\approx \left[A_x^{(0)}\right]^2+2A_x^{(0)}A_x^{(1)}+\left[\left[A_x^{(1)}\right]^2+2A_x^{(0)}A_x^{(2)}\right].
    \label{eq:numerator}
\end{eqnarray}
Combining Eqs.~(\ref{eq:numerator}) and (\ref{eq:denominator}), one obtains
\begin{eqnarray}
    F_{c,\psi}=\sum_{k=0}^\infty F_{c,\psi}^{(k)},
\end{eqnarray}
where 
\begin{eqnarray}
    F_{c,\psi}^{(1)}&=&\sum_n \delta \phi_n F_{c,n}^{(1)},\nonumber\\
    F_{c,\psi}^{(2)}&=&\sum_{n1,n_2}\delta\phi_{n_1}\delta\phi_{n_2}F_{c,n_1n_2}^{(2)},
\end{eqnarray}
with 
\begin{eqnarray}
F_{c,\psi}^{(0)}&=&\sum_x \frac{(A_x^{(0)})^2}{P_x^{(0)}},
\end{eqnarray}
being the CFI in the disorder-free case, and
\begin{eqnarray}
F_{c,\psi,n}^{(1)}&=& \sum_x  \left(\frac{2A_x^{(0)}A_{x,n}^{(1)}}{P_x^{(0)}}-\frac{(A_x^{(0)})^2 P_{x,n}^{(1)}}{(P_x^{(0)})^2}\right),\nonumber\\
F_{c,\psi,n_1n_2}^{(2)}&=&\sum_x\Bigg[(A_x^{(0)})^2\left(\frac{P_{x,n_1}^{(1)}P_{x,n_2}^{(1)}}{(P_x^{(0)})^3}-\frac{P_{x,n_1n_2}^{(2)}}{(P_x^{(0)})^2}\right)-\frac{2A_x^{(0)}A_{x,n_1}^{(1)}P_{x,n_2}}{(P_x^{(0)})^2}+\frac{A_{x,n_1}^{(1)}A_{x,n_2}^{(1)}+2A_{x,n_1n_2}^{(2)}A_{x}^{(0)}}{P_x^{(0)}}\Bigg].
\end{eqnarray}
We will be specifically interested in $F_{c,\psi,n_1n_2}^{(2)}$, which, with $n_1=n_2=n$, becomes
\begin{eqnarray}
\label{eq:Fc2}
 F_{c,\psi,n}^{(2)}&=&\sum_x\left[(A_x^{(0)})^2\left(\frac{(P_{x,n}^{(1)})^2}{(P_x^{(0)})^3}-\frac{P_{x,n}^{(2)}}{(P_x^{(0)})^2}\right)-\frac{2A_x^{(0)}A_{x,n}^{(1)}P_{x,n}^{(1)}}{(P_x^{(0)})^2}+\frac{(A_{x,n}^{(1)})^2+2A_{x,n}^{(2)}A_{x}^{(0)}}{P_x^{(0)}}\right].
\end{eqnarray}

\section{Calculation of \texorpdfstring{$C_{c,n}^{\psi,(2)}$}{Cc2} from the spectrum of clean-probe Hamiltonian}
\label{app:C_classical_calculation}

We work in the energy eigenbasis of the clean Hamiltonian introduced
in Appendix~\ref{app:C_calculation}, in which  a measurement operator $M_x$ can be represented as
\begin{equation}
\mathcal{M}_{x,t}
    =
    U_{0,t}^\dagger M_x U_{0,t}
    =
    S M_x'(t)S^\dagger,
\label{D1}
\end{equation}
where
\begin{equation}
\left[M_x^{'}(t)\right]_{ij}
=
\left[\tilde{M_x}\right]_{ij}
e^{\mathrm{i}(E_i-E_j)t}.
\label{D2}
\end{equation}
The zeroth-order contribution to the interaction-picture measurement
operator is then 
\begin{equation}
M_{x,t}^{(0)}
=
\mathcal{M}_{x,t},
\label{D3}
\end{equation}

For convenience, we introduce the following functions:
\begin{equation}
A_{kji} =
\begin{cases}
t\,e^{\mathrm{i}t\Delta E_{ij}}, & E_i=E_k,\\[6pt]
e^{\mathrm{i}t\Delta E_{kj}}T_{ik}, & E_i\neq E_k.
\end{cases}
\label{D4}
\end{equation}
and
\begin{equation}
B_{ilk}
=
\begin{cases}
\left(e^{\mathrm{i}t\Delta E_{il}}
\left[1-\mathrm{i}\Delta E_{il} t\right]-1\right)\Delta E_{il}^2,
& E_i=E_k,\quad E_i\neq E_l,\\[10pt]
\dfrac{t^2}{2},
& E_i=E_k=E_l,\\[6pt]
-\mathrm{i}(T_{il}-T_{kl})\Delta E_{ik},
& E_i\neq E_k.
\end{cases}
\label{D5}
\end{equation}

\begin{equation}
C_{iklj}
=
\begin{cases}
-
\mathrm{i} e^{\mathrm{i}t\Delta E_{il}}
\left(e^{\mathrm{i}t\Delta E_{lj}}-1\right)t\Delta E_{lj}
,
& E_i=E_k,\quad E_l\neq E_j,\\[10pt]

-\mathrm{i} e^{\mathrm{i}t\Delta E_{kj}}
\left(e^{\mathrm{i}t\Delta E_{ik}}-1\right)t\Delta E_{ik},
& E_i\neq E_k,\quad E_l=E_j,\\[10pt]

t^2 e^{\mathrm{i}t\Delta E_{kj}},
& E_i=E_k,\quad E_l=E_j,\\[8pt]

T_{ik}e^{\mathrm{i}t\Delta E_{kl}}T_{lj},
& E_i\neq E_k,\quad E_l\neq E_j.
\end{cases}
\label{D6}
\end{equation}
Here, $T_{ij}$ is as defined in Eq.~\eqref{eq:Tij}. Subsequently, the first-order contribution to the
measurement operator in the clean energy-eigenbasis is given by
\begin{equation}
 [M_{x,n}^{(1)}]_{ij}= \mathrm{i}\sum_k \left[\tilde{H_n}\right]_{ik}
    \left[\tilde{M_x}\right]_{kj} A_{kji}-\mathrm{i}\sum_k
    \left[\tilde{M_x}\right]_{ik}
    \left[\tilde{H_n}\right]_{kj}A^*_{kij}
    \label{D7}
\end{equation}
while the second-order contribution is
\begin{equation}
\begin{aligned}
  \left[M_{x,n}^{(2)}\right]_{ij}
    ={}&
    -\sum_{k,l}
    [\widetilde{H_n}]_{ik}
    [\widetilde{H_n}]_{kl}
    [\widetilde{M_x}]_{lj}
    e^{\mathrm{i}\Delta E_{lj}t} B_{ilk}-
    \sum_{k,l}
    [\widetilde{M_x}]_{ik}
    [\widetilde{H_n}]_{kl}
    [\widetilde{H_n}]_{lj}
    e^{\mathrm{i}\Delta E_{ik}t}B^*_{jkl}\\
    &+\sum_{k,l}(\widetilde{H_n})_{ik}
    [\widetilde{M_x}]_{kl}
    [\widetilde{H_n}]_{lj} C_{iklj}
    \end{aligned}.
    \label{D8}
\end{equation}
Using these along with Appendix~\ref{app:C_calculation}, $F_{c,\psi,n}^{(2)}$ and hence $C_{c,n}^{\psi,(2)}$
can be evaluated directly from the clean Hamiltonian, the initial
probe state, and the chosen measurement operator, without a disorder averaging.

\section{Proof of Eq.~(\ref{eq:lambda_12})}
\label{app:proof_of_hierarchy}

We start with $D(\varepsilon)=\lambda_1\varepsilon+\lambda_2\varepsilon^2+D^\prime(\varepsilon)$,  
where $\lambda_1=4\tilde{G}_n^\psi-F_{c,\psi,n}^{(1)}$, $\lambda_2=4\tilde{G}_n^{\psi,(2)}-F_{c,\psi,n}^{(2)}$, and  $D^\prime(\varepsilon)\sim\mathcal{O}(|\varepsilon|^3)$. For a function $D(\varepsilon)$ that is smooth near $\varepsilon=0$, by Taylor's theorem, two constants $\varepsilon_0\geq 0$ and $K\geq 0$ exist such that $|D^\prime(\varepsilon)|\leq K|\varepsilon|^3$ for $|\varepsilon|\leq \varepsilon_0$. Further, discussions following Eq.~(\ref{eq:difference}) imply $D(\varepsilon)\geq0\forall\varepsilon$ and $D(0)=0$.  Assuming $\lambda_1\neq0$, and taking $\varepsilon=-\mathrm{sgn}(\lambda_1)\eta$ with $0<\eta\leq\varepsilon_0$ such that $\lambda_1\varepsilon=-|\lambda_1|\eta$,
\begin{eqnarray}
D(\varepsilon)\leq-|\lambda_1|\eta+|\lambda_2|\eta^2+K\eta^3=-\eta\left[|\lambda_1|-|\lambda_2|\eta-K\eta^2\right].
\end{eqnarray}
Since $|\lambda_2|\eta+K\eta^2\to0$ as $\eta\to0$, we choose $\eta$ small enough such that $|\lambda_2|\eta+K\eta^2<|\lambda_1|$. This leads to the term in the square bracket to be $+$ve, $D(\varepsilon)<0$, which is a contradiction. Therefore, $\lambda_1=0$. This further leads to $D(\varepsilon)=\lambda_2\varepsilon^2+D^\prime(\varepsilon)$. Next, we assume $\lambda_2<0$. For $0<|\varepsilon|\leq\varepsilon_0$,
\begin{eqnarray}
D(\varepsilon)\leq-|\lambda_2|\varepsilon^2+K|\varepsilon|^3=-\varepsilon^2\left[|\lambda_2|-K|\varepsilon|\right],
\end{eqnarray}
which is $-$ve for $0<|\varepsilon|<\min\{\varepsilon_0,|\lambda_2|/K\}$ (any $0<|\varepsilon|\leq\varepsilon_0$ if $K=0$), indicating another contradiction. Hence $\lambda_2\geq0$.

\twocolumngrid 

\bibliography{ref}

\end{document}